\documentclass[twocolumn,epjc3]{svjour3}  
\journalname{Eur. Phys. J. C}

\usepackage{graphicx}%
\usepackage{multirow}%
\usepackage{amsmath,amssymb,amsfonts}
\usepackage{mathrsfs}%
\usepackage{footmisc}
\usepackage{xcolor}%
\usepackage{txfonts}
\usepackage[T1]{fontenc}
\usepackage[square, numbers]{natbib}
\usepackage{bigdelim}

\definecolor{light-gray}{gray}{0.95}
\definecolor{dark-gray}{gray}{0.4}

\definecolor{light-gray}{gray}{0.95}
\definecolor{dark-gray}{gray}{0.4}

\newcommand{\ajax}{{AJAX}}
\newcommand{\apache}{{APACHE}}
\newcommand{\crdb}{{\tt CRDB}}
\newcommand{\crdbv}[1]{{\tt CRDB~\!v{#1}}}

\newcommand{\kcdc}{{\tt KCDC}}

\newcommand{\javascript}{{JavaScript}}
\newcommand{\jquery}[1]{{\tt jquery{#1}}}
\newcommand{\lamp}{{LAMP}}
\newcommand{\linux}{{Linux}}
\newcommand{\mysql}{{MySQL}}
\newcommand{\php}{{PHP}}
\newcommand{\rest}{{REST}}

\newcommand{\tsorter}{{\tt table-sorter}}

\newcommand{\xml}{{XML}}

\newcommand{\code}[1]{\texttt{#1}}

\graphicspath{figures/} 

\usepackage{hyperref}
\hypersetup{breaklinks=true,%
            colorlinks=true,%
            citecolor=blue,
            linkcolor   =red,
            anchorcolor =gray,
            filecolor   =cyan,
            menucolor   =red,
            runcolor    =cyan,
            urlcolor    =magenta,
            pdfauthor={Maurin, Ahlers, Dembinski et al.},%
            pdftitle={CRDB v4.1}%
           }

\begin{document}

\title{A cosmic-ray database update\thanks{\href{mailto:crdb@lpsc.in2p3.fr}{crdb@lpsc.in2p3.fr}}: CRDB v4.1}
\titlerunning{\crdbv{4.1}}

\author{
        David~Maurin$^1$ \and
        Markus~Ahlers$^2$ \and
        Hans~Dembinski$^3$ \and
        Andreas~Haungs$^4$ \and
        Pierre-Simon~Mangeard$^5$ \and
        Frédéric~Melot$^1$ \and
        Philipp~Mertsch$^6$ \and
        Doris~Wochele$^4$ \and
        Jürgen~Wochele$^4$
    }
\authorrunning{D. Maurin, M. Ahlers, H.  et al.}

\institute{
LPSC, Universit\'e Grenoble-Alpes, CNRS/IN2P3, 53 avenue des Martyrs, 38026 Grenoble, France
\and
Niels Bohr Institute, University of Copenhagen, Blegdamsvej 17, DK-2100 Copenhagen, Denmark
\and
Department of Physics, TU Dortmund, Otto-Hahn-Straße 4a, 44227, Dortmund, Germany
\and
Institute for Astroparticle Physics (IAP), Karlsruhe Institute of Technology (KIT), Karlsruhe, Germany
\and
Bartol Research Institute, University of Delaware, Newark, DE 19716, USA
\and
Institute for Theoretical Particle Physics and Cosmology (TTK),
RWTH Aachen University, 52056 Aachen, Germany
}

\date{Received: date / Accepted: date}

\maketitle

\begin{abstract}
The cosmic-ray database, \crdb{}, has been gathering cosmic-ray data for the community since 2013.
We present a new release, \crdbv{4.1}, providing many new quantities and data sets, with several improvements made on the code and web interface, and with new visualisation tools.
\crdb{} relies on the \mysql{} database management system, \jquery{} and \tsorter{} libraries for queries and sorting, and \php{} web pages and \ajax{} protocol for displays. A \rest{} interface enables user queries from command line or scripts. A new (pip-installable) \crdb{} python library is developed and extensive jupyter notebook examples are provided.
This release contains cosmic-ray dipole anisotropy data, high-energy $\bar{p}/p$ upper limits, some unpublished LEE and AESOP lepton time series, many more ultra-high energy data, and a few missing old data sets. It also includes high-precision data from the last three years, in particular the hundreds of thousands AMS-02 and PAMELA data time series (time-dependent plots are now enabled). All these data are shown in a gallery of plots, which can be easily reproduced from the public notebook examples.
\crdb{} contains 316$\,$126 data points from 504 publications, in 4111 sub-experiments from 131 experiments.
\end{abstract}

\section{Introduction}

Owing to the quantity and variety of data gathered in cosmic-ray (CR) physics, a central shared database (DB) assuring data quality, completeness, and traceability is an asset for the community. Although the oldest datasets have a historical value mostly, the low-energy data still trace and give a unique perspective on the 11-year Solar cycle \cite[e.g.][]{2017AdSpR..60..833G,2019ApJ...887..132S}, and may also be of unforeseen use in the future.

The Cosmic-Ray DataBase\footnote{\url{https://lpsc.in2p3.fr/crdb}} (\crdb{}) team has been distributing a growing body of CR data since its first public release in 2013 \cite{2014A&A...569A..32M}. In a recent update, \crdbv{4.0} \cite{2020Univ....6..102M}, existing data on (groups of) ultra-heavy elements ($Z>30$), upper limits on anti-nuclei ($Z\leq-2$), and a selected sample of ultra-high-energy (UHE) CRs from ground-experiments were included. In \crdbv{4.0}, the DB structure and the submission data format were also revised, and users were provided with a \rest{} interface to extract both CR data and solar modulation levels (in their own codes and scripts), with overall more flexibility and more keywords to select the data queried.

In this release, \crdbv{4.1}, beside uploading data from the last three years (from AMS-02, CALET, DAMPE, PA\-MELA, etc.), we take advantage of an agreement with our colleagues from the \kcdc{}\footnote{\url{https://kcdc.iap.kit.edu/}} DB \cite{2018EPJC...78..741H} to complete our sample of UHECR data. We also add energy-dependent anisotropy data, including and extending those presented in \cite{2017PrPNP..94..184A}. We also correct the meta-data and provide a few unpublished low-energy leptons and positron fraction data from the LEE, AESOP and AESOP-LITE balloon flights (operated over a 50 year time period). Because an incredibly large body of time-dependent data has been released by the AMS-02 experiment, we provide a new interface to ease the visualisation of these time series; these data are now the most numerous by far in \crdb{}. One of the main novelty of this release is a new standalone python library for the plotting of \crdb{} data, which should further ease their distribution and use by the community at large. We also took the opportunity of this release to fix some mistakes in the data, meta-data, and to improve the code (behind the scene) and the web interface; the most important changes are documented and available on \crdb{}'s webpage, and briefly described later on.

The paper is organised as follows: Sect.~\ref{sec:db_content} recalls the DB structure and the few changes made in this release; Sect.~\ref{sec:web_queries} presents the web interface and its novelties, and also introduce the new public python library to query and display \crdb{} data (outside of the website); Sect.~\ref{sec:v4.1-newdata} highlights the new data added in this version; we conclude in Sect.~\ref{sec:concl}.
\ref{app:Eunit} motivates our new (and hopefully more rationale) convention for the energy units in \crdb{}.\textbf{}

\begin{figure*}
\begin{center}
\includegraphics[width=0.9\textwidth]{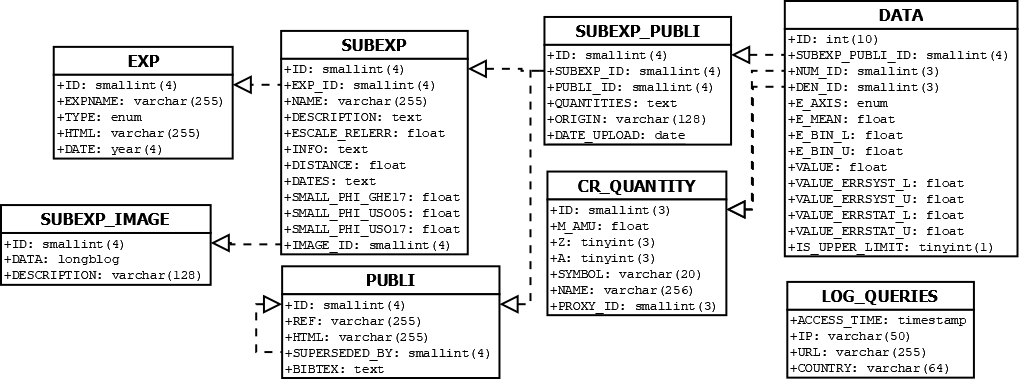}
\caption{Tables and keys in the \mysql{} structure of \crdb{}. The data (energy, values, and uncertainties) are stored in \code{DATA} and \code{CR\_QUANTITY} tables. The meta-data (publication, experiment and sub-experiment names and infos) are stored in the \code{EXP}, \code{SUBEXP}, \code{SUBEXP\_IMAGE}, and \code{PUBLI} tables, with \code{SUBEXP\_PUBLI} a bridge table enabling to access and link these various meta-data. The \code{ISOTOPE\_PROXY} table is used to define the rules for energy-axis conversions of CR fluxes (see App.~A.4 of \cite{2020Univ....6..102M} or in the new `Caveats/Tips' web page, see Sect.~\ref {sec:webinterface}). The  \code{LOG\_QUERIES} table keeps track of the number and origin of the visits.}
\label{fig:mysql}
\end{center}
\end{figure*}

\section{Database structure}
\label{sec:db_content}

In \crdb{}, data are separated in two broad categories, namely the {\em data} (CR data points and data uncertainties) and the {\em meta-data} (data about the data): the latter include the data taking periods, the description of the experiment, links to the associated publications, etc.
The DB structure, shown in Fig.~\ref{fig:mysql}, has only slightly changed since our last release. Its most important features are recalled below, and we use \code{MONOSPACE} font to easily identify the DB table names and keys.

\subsection{Data points and energy axis (\code{DATA} table)}
\label{sec:table-data}
Data points are described in the \code{DATA} table (see Fig.~\ref{fig:mysql}). Each entry has a unique \code{ID} and corresponds to a measured \code{VALUE} or upper limit (if boolean \code{IS\_UPPER\_LIMIT} set to 1) within an energy bin [\code{E\_BIN\_L}, \code{E\_BIN\_U}] or at the mean energy bin value \code{E\_MEAN}\footnote{If only \code{E\_MEAN} is provided in the publication, we set \code{E\_BIN\_L}~$=$~\code{E\_BIN\_U}~$=$~\code{E\_MEAN}. If both \code{E\_BIN\_L} and \code{E\_BIN\_U} are provided but not \code{E\_MEAN}, we set $\code{E\_MEAN}=(\code{E\_BIN\_L}\times\code{E\_BIN\_U})^{1/2}$. Finally, some experiments define their last energy bin as all events above a given energy: in that case, we manually set an upper bin value at least 100 times the lower bin value.}. The data point is also associated to a sub-experiment and publication via its \code{SUBEXP\_PUBLI\_ID} key (whose value points at a \code{SUBEXP\_PUBLI} table entry, see Sect.~\ref{sec:table-subexppubli}).

To cover the different energy types provided in the original publications, the energy axis (\code{E-AXIS}) of each data point must be set to \code{ETOT}, \code{EK}, \code{R}, \code{EKN}, or \code{ETOTN}. These types correspond to and are given in unit of, respectively, total energy $E_{\rm tot}$ in GeV, kinetic energy $E_{\rm k}\!=\!E_{\rm tot}-m$ in GeV, rigidity ${\cal R}\!=\!pc/(Ze)$ in GV, kinetic energy per nucleon $E_{\rm k/n}\!=\!E_{\rm k}/A$ in GeV, and total energy per nucleon $E_{\rm tot/n}\!=\!E_{\rm tot}/A$ in GeV: as discussed in \ref{app:Eunit}, we changed the energy unit convention of $E_{\rm k/n}$ from GeV/n into GeV in this version. For the data, \crdb{} enables asymmetric statistical (\code{VALUE\_ERRSTAT\_L} and \code{VALUE\_ERRSTAT\_U}) and systematic (\code{VALUE\_ERRSYST\_L} and \code{VALUE\_ERRSYST\_U}) uncertainties\footnote{For old data, the distinction was usually not made between the two, and because old measurements were mostly limited by their statistics, the quoted uncertainties in the publications are ascribed to \code{VALUE\_ERRSTAT\_L} and \code{VALUE\_ERRSTAT\_U}.}.

\subsection{Quantities and conversions (\code{CR\_QUANTITY} table)}
\label{sec:table-qties}
The measured quantity is either a single CR quantity \code{NUM\_ID} or a ratio of two CR quantities \code{NUM\_ID}/\code{DEN\_ID}, where both \code{NUM\_ID} and \code{DEN\_ID} point to entries in the \code{CR\_QUANTITY} table. These entries are identified by an \code{ID} (set manually), a \code{SYMBOL}, and a \code{NAME}. The keys \code{A}, \code{Z}, and \code{M\_AMU} (for the atomic mass number, charge, and mass in a.m.u) are non-null for isotopes, only the key \code{Z} can be filled for elements, and all keys are set to zero for groups of elements (or compound quantities) and dipole anisotropy data.

In \crdb{} queries, the data conversion from one energy axis to another is enabled (see Table~A.1 in \cite{2020Univ....6..102M}). The conversion is exact for individual fluxes of CR isotopes or leptons and for ratios of leptons, and also for $\bar{p}/p$ (this last conversion was not implemented in the previous release), but it is impossible for generic ratios, compound quantities, or anisotropy data. Nevertheless, an approximate conversion can still be enforced for fluxes of elements (or group of elements) if these quantities have a CR isotope proxy; this proxy is enabled via the \code{PROXY\_ID} key in the \code{CR\_QUANTITY} table (this key was previously in a separate and redundant table that we removed in this release).

\subsection{Meta-data for experiments and modulation level (\code{EXP}, \code{SUBEXP}, and \code{SUBEXP\_IMAGE} tables)}
\label{sec:tables-expsubexp}

\paragraph{Definition and description.}
CR data are taken from experiments described in the \code{EXP} table (see Fig.~\ref{fig:mysql}). Each experiment has a \code{TYPE} (\code{balloon}, \code{ground}, or \code{space}), a unique \code{ID} (set internally in the DB), a name (\code{EXPNAME}), a starting year (\code{DATE}), and optionally a website (\code{HTML}); we stress that the experiment name is mainly used to better regroup and sort sub-experiments in the {\em Experiments/Data} website tab.

Sub-experiments (\code{SUBEXP} table) have an \code{ID} and are attached to a single experiment (\code{EXP\_ID}). They enable to tag and distinguish, for a same experiment: (i) data obtained from different data taking periods; (ii) data taken from distinct sub-detectors or reconstructed from different analysis types; (iii) data obtained using external third-party models or different assumptions. Sub-experiments have a \code{NAME}\footnote{In \crdb{}, we decided that the format of this name should be a concatenation of: (i) the experiment name \code{EXPNAME} (e.g. {\em PAMELA}, {\em NUCLEON}); (ii) if necessary, a hyphen-separated sub-detector characteristic (e.g., {\em PAMELA-CALO}) or specific technique used (e.g. {\em NUCLEON-KLEM}); (iii) data taking periods in parenthesis; (iv) if relevant, the Monte Carlo generator used to analyse the data (e.g. {\em IceTop SIBYLL2.1} for UHECR data).
The two exceptions to form the sub-experiment names are for the case of a combined analysis, i.e. names based on the concatenation of the two experiment names, e.g. {\em IceCube+IceTop (2010/06-2013/05)}, and unnamed balloons (concatenation of {\em Balloon} and their flight dates).
For the dates, the chosen format is {\em YYYY/MM} (shortened to {\em YYYY} if the month is unknown), with a single date for a shorter-than-a-month data taking period, e.g. {\em Balloon (1966/05)}, or two dates otherwise, e.g. {\em IMP7 (1973/05-1973/08)}; if the month is unknown, we only quote the year (or range of years).}, a short \code{DESCRIPTION} (detector or detection technique), additional \code{INFO} (e.g. location for balloon flights, GPS coordinates for ground-based detectors, etc.), and an \code{IMAGE\_ID} (see next). For each sub-experiment, we also provide a single value (set to zero by default) for a possible energy-scale relative uncertainty (\code{ESCALE\_RELERR}).

In this release, we also added the new \code{SUBEXP\_IMAGE} table (see Fig.~\ref{fig:mysql}). Previously, the detector images were kept in a separate directory with file names based on the \code{EXPNAME} and or sub-experiment \code{NAME} keys. In the new table, we have the image itself (\code{DATA} key) with its unique \code{ID} key, along with a brief description if needed (\code{DESCRIPTION} key). This allows to avoid storing duplicate images and makes checks on the completeness of the presence of images for all sub-experiments easier.

\paragraph{Solar modulation level.}
Especially important for the interpretation of low-energy data (below a few hundreds of GeV), we must provide (i) the \code{DISTANCE} to the Sun of the sub-experiment---almost all experiments are at 1~a.u., but a few satellites ({\em Ulysses} and {\em Voyager}) have also taken data at different position inside and outside the Solar cavity--- and (ii) the exact list of start-stop \code{DATES} of the data taking periods\footnote{The format is \code{YYYY/MM/DD:HHMMSS-YYY/MM/DD:HHMMSS}, or a semi-column separated list of similarly formatted time periods if necessary. If the exact time is unavailable, we enforce \code{HHMMSS=000000} for the start date and \code{HHMMSS=235959} for the stop date. If the day is unknown, we enforce \code{DD=01} (start) or last day of the month (stop), and if the month is unknown, we enforce \code{MM=01} (start) and \code{MM=12} (stop).}. These two pieces of information allow to calculate and fill \code{SMALL\_PHI}, the average modulation level over the corresponding data taking periods, in the force-field approximation \cite{1967ApJ...149L.115G,1968ApJ...154.1011G}. Actually, \code{SMALL\_PHI} contains different estimates of $\langle\phi(t)\rangle$, all calculated from the same neutron monitor data\footnote{\url{http://www01.nmdb.eu}}, but based on slightly different modellings: the values tagged [Uso05] and [Uso17] are based on monthly average public values\footnote{\url{http://cosmicrays.oulu.fi/phi}} from \cite{2005JGRA..11012108U,2017JGRA..122.3875U}, while those tagged [Ghe17] are based on daily average values from \cite{2017AdSpR..60..833G}. In \crdb{}, all queried data are returned with their calculated  \code{SMALL\_PHI} value, but users are obviously free to discard or re-calculate it---by default, the returned values are [Ghe17], which can be also calculated  for any time period from the {\em Solar modulation} tab (see Sect.~\ref{sec:webinterface}).

\subsection{Meta-data for publications (\code{PUBLI} table)}
\label{sec:table-publi}

Almost all data in \crdb{} are taken from peer-reviewed publications. The main exceptions are data from balloon flights before the 1990's, which were published in the proceedings of the biennial International Cosmic-Ray Conference only. Each publication is stored in the \code{PUBLI} table (see Fig.~\ref{fig:mysql}) with  a unique \code{ID} (set internally) and an  \code{HTML} key, taken to be the publication ADS (Astrophysics Data System)  identifier (e.g. \href{https://ui.adsabs.harvard.edu/abs/2014A&A...569A..32M}{\code{2014A\&A...569A..32M}}). This identifier allows to retrieve and fill in a standardised manner the \code{REF} and \code{BIBTEX} keys via the ADS API\footnote{\url{https://github.com/adsabs/adsabs-dev-api}}. The original publications are stored in \crdb{} (for the administrators) but cannot be made publicly available because of publication rights.

Because some data sets are sometimes re-analysed and reported in a new publication, we set the \code{SUPERSEDED\_BY} key of the obsolete one to the \code{ID} value of the new one (it is left empty if it is not superseded). This allows us to enforce that queries to \crdb{} always return the most recent data, discarding the deprecated ones. We nevertheless keep track of these superseded data in the `Experiments/Data' tab (see Sect.~\ref{sec:webinterface}), where old and new publications are shown.

\subsection{Tying data and meta-data (\code{SUBEXP\_PUBLI} table)}
\label{sec:table-subexppubli}

The full description of the data requires the data themselves, the sub-experiment that measured them, and the publication where they appeared. The \code{SUBEXP\_PUBLI} bridge table (see Fig.~\ref{fig:mysql}) allows to tackle situations where several sub-experiments are reported in the same publication. Each data set with a unique \code{ID} is tied to a sub-experiment (\code{SUBEXP\_ID}) and a publication (\code{PUBLI\_ID}).
In addition, in this table, we keep track of the date at which each dataset was uploaded in \crdb{} (\code{DATE\_UPLOAD}), and also of all CR \code{QUANTITIES} whose data were provided in this publication. While both these keys are unused in data queries, they are useful for maintenance and cross-checks of the DB.

\section{Web interface and queries}
\label{sec:web_queries}

\crdb{} runs on free open source softwares with a classical \lamp{} solution: \linux{} operating system, \apache{} HTTP server, \mysql{} database, and \php{} scripting language.
The server is hosted at the LPSC laboratory, and has been recently changed to have a more recent version of the operating system, the DB, and the \php{} version. The DB RAM was extended from 512~MB to 2048~MB to handle the larger requests from the newly added time-series data (see Sect.~\ref{sec:v4.1-time-series}).
The \crdb{} website is organised in tabs providing different entry points to explore the DB data and meta-data. The webpages use \ajax{} (asynchronous \javascript{} and \xml{}) web development technique for efficiency and speed. In addition to the few improvements made on the existing website tabs, we added two new ones in this release (see Sect.~\ref{sec:webinterface}).
To query, sort, and show the DB content, the web interface relies on \jquery{}, \jquery{-ui}, \jquery{.cluetip}, and \tsorter{}. There are two ways for users to query data: either from the {\em Data extraction} tab (see below) or from a direct command-line call (bypassing the website) via the \rest{} interface (also see below). The latter functionality has been fully exploited in this release, with the development of a new dedicated \crdb{} python library. This library is described and used to generate a gallery of plots in Sect.~\ref{sec:pycrdb}.

\subsection{Web pages: content and novelties}
\label{sec:webinterface}

We briefly describe below the content and noteworthy improvements made on the tabs. We also added a new tab to list a few caveats and tips related to the data preparation and transformations.

\begin{itemize}
  \item {\em Welcome} tab: entry point of the website, where the DB content, tools, people involved, code status, etc. are highlighted. In this release, we also added a gallery of plots to advertise the variety of data in \crdb{}.

   \item {\em Caveats/Tips} tab: there are a few subtleties in the way the data (and meta-data) are handled in \crdb{}. Indeed, at the collection stage, the information on the data is sometimes partial, and somewhat subjective choices need to be made to be able to implement them nonetheless. Then, at the query stage, combinations and conversions are enabled, with some degree of approximation as well. Users probably do no pay a lot of attention to these details, and this is probably fine most of the time. Whereas the details and caveats about these procedures are made explicit in the \crdb{} publications \cite{2014A&A...569A..32M,2020Univ....6..102M}, the most relevant ones are gathered here in one place. This should help users identify data for which going back to the original publication is necessary.

  \item {\em Data extraction} tab: queries of user-selected CR quantities with various options (sub-experiment names, dates, energy unit, etc.). The retrieved data include the ones matching exactly the query but also, if selected, extra sets based on energy conversions (Table~A.1 of \cite{2020Univ....6..102M}) and data combinations (App.~A of \cite{2014A&A...569A..32M}); we added in this release the trivial but forgotten transformation rule to get Y/X from data published as X/Y. The data retrieved are then plotted and listed in a pop-up window and can be downloaded in various formats: in this release we added an extra option, `csv (as import)', enabling to retrieve the data and all their meta-data (format similar to the one described in the {\em Submit data} tab, see below). We also added a tick box for the `Refine search criteria' box in the {\em Data extraction} tab, to display the data versus time instead of energy.

   \item {\em Experiments/Data} tab: sorted list of experiments with their associated sub-experiments, including in particular a picture of the detector, their associated publications and quantities measured. In this release, to improve the sorting and readability of the numerous unnamed balloon flight series (i.e. balloon launched multiple times over years by the same team and analysed in several publications), we regrouped them into fewer and more informative names, e.g. {\em Nuclear emulsions 1950-1968}, {\em Muon Telescope 1957-1995}, etc.

   \item {\em REST/CRDB.py} tab: details how to query \crdb{} from a stand-alone script, with the same options as the ones provided in the {\em Data extraction} tab (datasets retrieved from the website or from the \rest{} interface with the same selection and options are the same). We also provide a simple command-line example (to run in a terminal) using {\tt curl}. This capability is taken advantage of and extended in this release thanks to a new standalone python library to retrieve and display data, for instance from a python notebook, see Sect.~\ref{sec:pycrdb}).

   \item {\em Solar modulation} tab: gives access, for any time interval, to the force-field modulation level (see Sect.~\ref{sec:tables-expsubexp}). Behind the scene, a {\em cron} scheduler downloads NM data daily from NMDB\footnote{\url{http://www01.nmdb.eu}\label{foot:nmdb}}. It also calculates the associated $\phi_{\rm FF}$, whose values can be retrieved for a selected time period and resolution (from 10~minute up to a month), either directly from this tab, or from a \rest{} interface. In this release, we fixed several minor bugs (as listed on the website), and more importantly, we fixed the broken REST interface and the daily update\footnote{All missing $\phi_{\rm FF}$ values were completed, and we also recalculated modulation levels, starting from 2015, for the THULE station (because of updated NM values in NMDB) and ROME station (using the correct number of NM tubes, which changed in 2017).}.

   \item {\em Submit data} tab: how to format and send a \code{csv} file to \crdb{}.

   \item {\em Useful links} tab: online resources related to CR data.

   \item {\em Admin} tab: maintenance tools to check broken or inconsistent entries and missing meta-data, detailed procedure to upload data in the DB. This tab is restricted to authenticated users (i.e. \crdb{} maintainers).
\end{itemize}

\subsection{Python access to CRDB (and notebook)}
\label{sec:pycrdb}

The \crdb{} provides a REST interface, which can be used from any programming language to automate downloading and processing data in scripts and programs. A tutorial on how to do this is available\footnote{\url{https://github.com/crdb-project/tutorial}}. Since Python is the dominant scripting language for data processing, we further provide a ready-made solution for Python users that simplifies and standardises queries from scripts. Users of this library do not need to learn the REST API, this is done internally by the library. The corresponding Python package called \code{crdb}\footnote{\url{https://github.com/crdb-project/crdb}} can be downloaded with the standard tool \code{pip} from the Python Package Index\footnote{\url{https://pypi.org/project/crdb}}. The main function is \code{crdb.query}, which performs a query to the database through keyword arguments, which are internally validated so that user errors are caught early and clear error messages are returned. The tabular output of a query is transformed by this function into a structured Numpy array \cite{harris2020array}, which allows for efficient fast processing in Python. Each query is automatically cached to disk for 30 days, to accelerate repeated calls to \code{crdb.query} and to reduce the load on the server; this often occurs during the development of a script or program.
Further utility functions allow users to easily generate lists of citations for the data sets they queried from the DB. All functions are well documented, the documentation can be accessed with Python's internal \code{help()} command.

\begin{figure*}
\centering
\includegraphics[width=\linewidth]{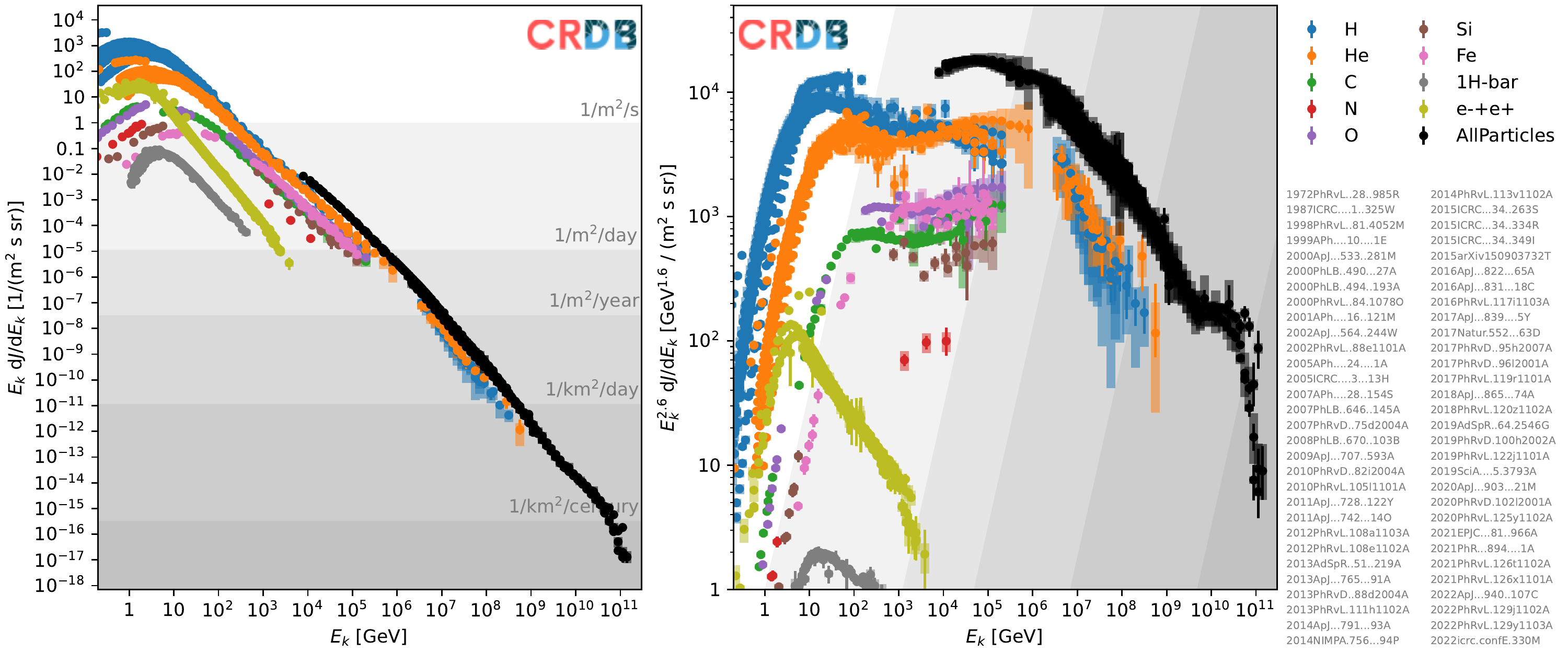}
\includegraphics[width=0.9\linewidth]{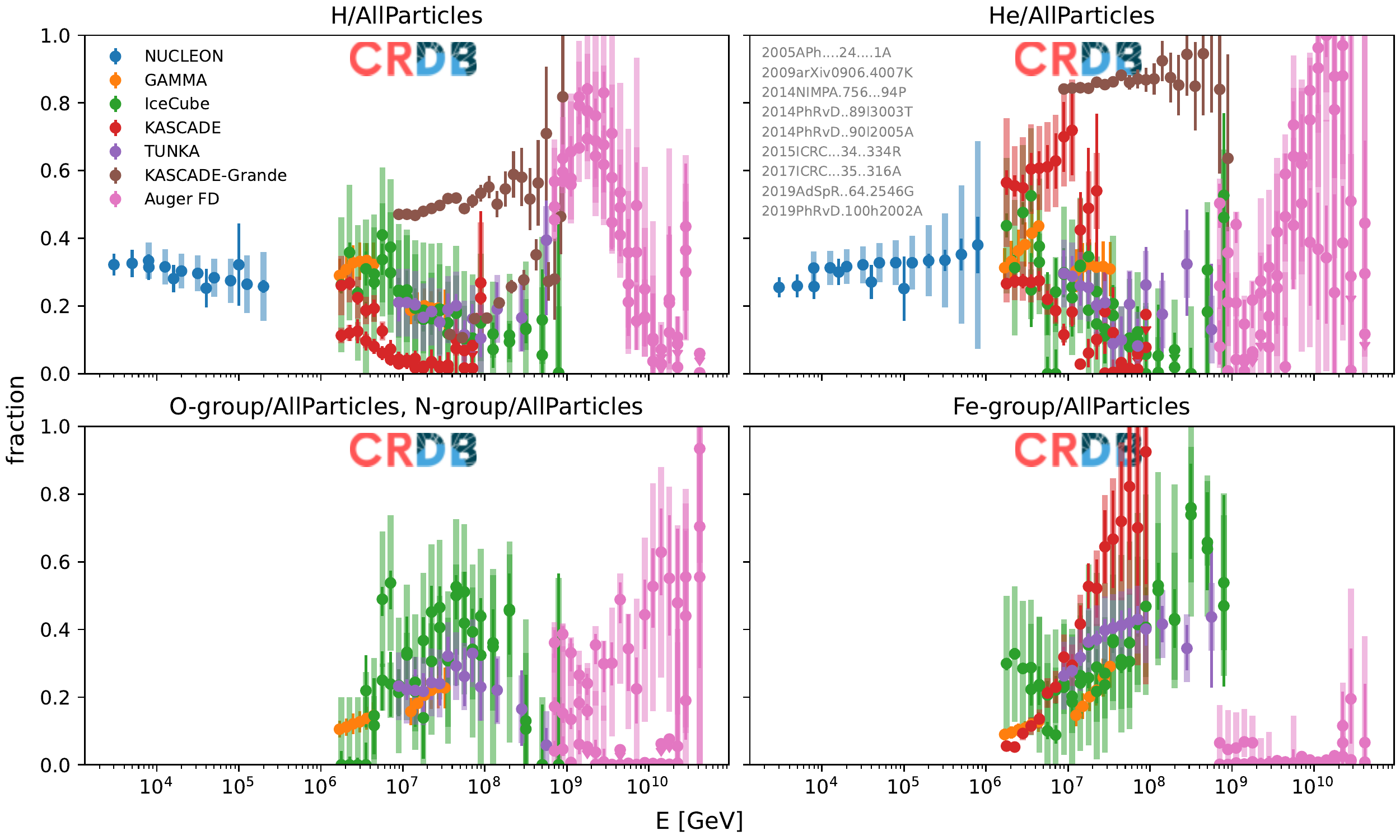}
\caption[]{Selected plots from the gallery, obtained from the \crdb{} python library and available from the gallery notebook\footref{foot:gallery}: flux of selected species (top), multiplied by $E_{\rm k}^{2.6}$ on the top right panel; energy dependence of high-energy CR groups of elements (bottom).}\label{fig:gallery_1}
\end{figure*}

\begin{figure*}
\centering
\includegraphics[width=\linewidth]{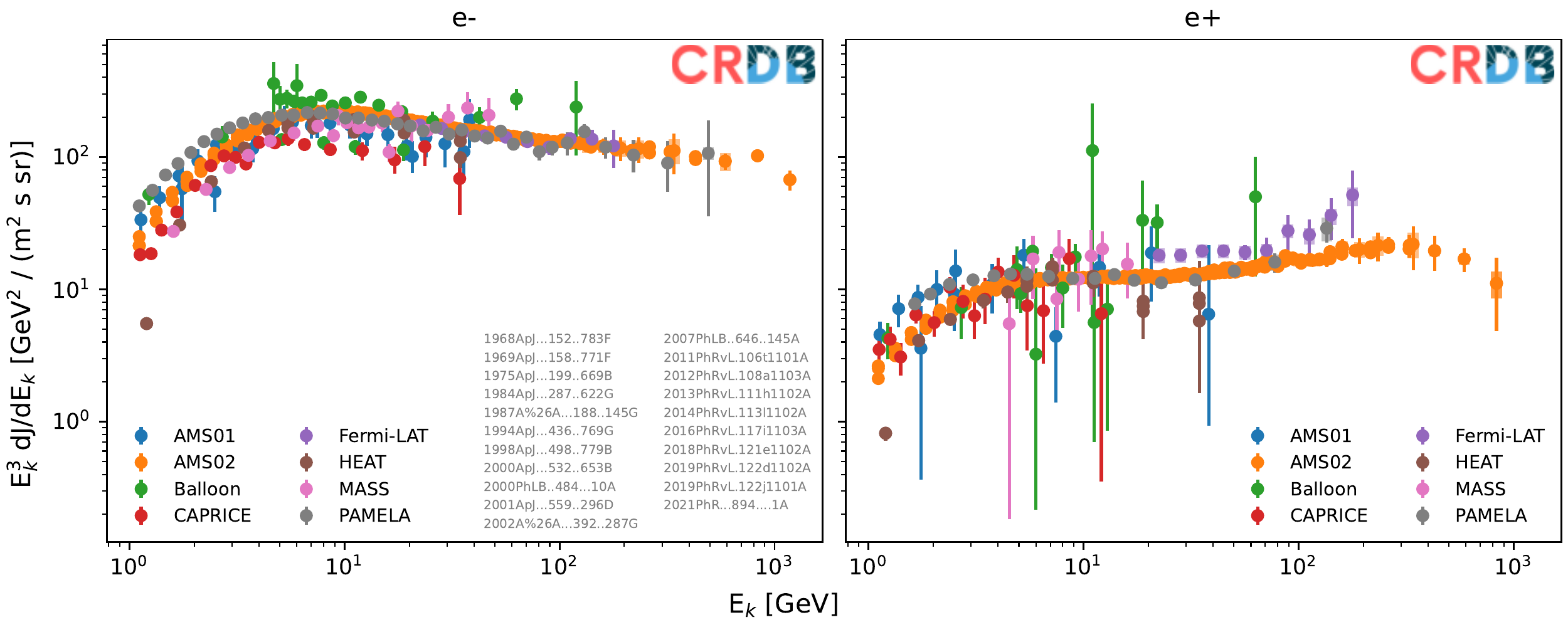}
\includegraphics[width=\linewidth]{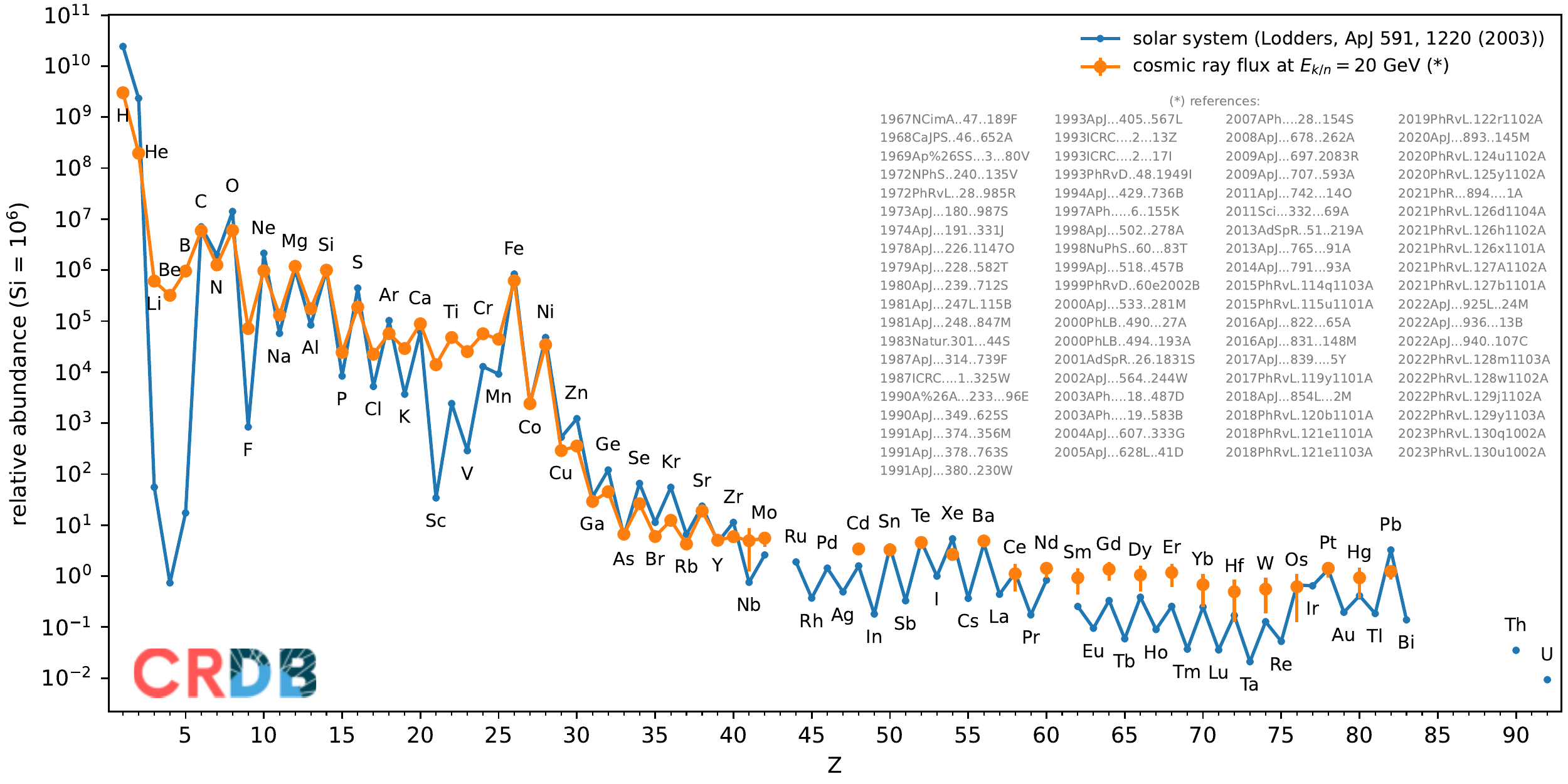}
\vspace{-0.5cm}
\caption[]{Selected plots from the gallery, obtained from the \crdb{} python library and available from the gallery notebook\footref{foot:gallery}:\textbf{} electron and positron fluxes (top), and comparison of elemental abundances in Solar system and GCRs (bottom).}
\label{fig:gallery_2}
\vspace{-0.2cm}
\end{figure*}

The Python package also provides a command-line interface, which allows users to perform queries and store the results in one of the ASCII formats supported by the \crdb{} data extraction system. In this case, the query is specified using command-line arguments, the latter mirroring those of \code{crdb.query}. Example code on how to make standard plots in Python can be found in the gallery, and we show in Figs.~\ref{fig:gallery_1} and~\ref{fig:gallery_2} a few plots illustrating the variety, coverage, and completeness of \crdb{}'s data. More plots are shown in the next section, and all of them are available from \crdb{}'s public gallery notebook\footnote{\url{https://github.com/crdb-project/tutorial/blob/main/gallery.ipynb}\label{foot:gallery}}.


\section{New datasets in \crdbv{4.1}}
\label{sec:v4.1-newdata}

In addition to {\em regular} data updated since the last release (Sect.~\ref{sec:v4.1-Std}), the content of \crdb{} has evolved in several directions. In this release, we (i) add dipolar anisotropy data (Sect.~\ref{sec:v4.1-Anisotropy}); (ii) take advantage of a partnership with \kcdc{} to gradually move from limited sample to completeness of UHECR data (Sect.~\ref{sec:v4.1-UHECRs}); (iii) include high-energy upper limits on antiproton fluxes from ground experiments ({Sect.~\ref{sec:v4.1-ULpbarp});  (iv) correct and complete low-energy lepton data from the LEE, AESOP, and AESOP-Lite balloons flown over 50 years  (Sect.~\ref{sec:v4.1-LEE-AESOP}); (v) expand time series data thanks to the recently released AMS-02 daily and PAMELA monthly data (Sect.~\ref{sec:v4.1-time-series}).

\begin{table*}
\caption{List of `regular' datasets added in the last three years, sorted according to their experiment \code{TYPE}: {\em ballon} (from Balloon to SOKOL), {\em ground} (OLIMPIYA only), and {\em space} (from ACE-CRIS to Voyager2).}
\label{table:new_generic}
\centering
{
\begin{tabular}{l l l r}
\hline\hline
  Subexp Name & Ref. & Qty & $N_{\rm data}$\\
\hline
  Balloon (1958/08b)            & \cite{1962NCim...26.1292F}\!\!\!\!\!& He                            & 7 \\
  Balloon (1974)                & \cite{1977ApL....18..125W} & $^{6,7}$Li/Li, $^{7,9,10}$Be/Be, $^{11}$B/B& 14\\
  Balloon (1975/10)             & \cite{1979ApJ...228..582T} & Co/Fe, Cu/Fe                  & 2 \\
  Balloon (1975/10+1976/05)     & \cite{1979ApJ...228..582T} & Ni/Fe, Zn/Fe                  & 2 \\
  Balloon (1977/09)             & \cite{1979ICRC....1..389W}$^{(a)}$ & $^{7}$Li/$^{6}$Li,\; $^{9}$Be/$^{7}$Be,\; $^{11}$B/$^{10}$B & 6 \\
  Balloon (1981/12)             & \cite{1983Natur.301...44S} &  Fe & 1 \\
Balloon-EWAsym (1989/05+1991/05) \hspace{0.57cm}\rdelim\}{2}{-2mm}& \multirow{2}{*}{\cite{1997APh.....6..155K}$^{(b)}$}  & \multirow{2}{*}{Si/Fe, S/Fe, Ar/Fe, Ca/Fe, (Z=21-25)/Fe} & \multirow{2}{*}{165}\\
Balloon-OpeningAngle (1989/05+1991/05) & & &\\
 BACH (1998/04)                & \cite{2002APh....16..387C} &  Fe & 1 \\
 HEIST (1984/05)            & \cite{1990ICRC....3...53G}  & $^{54}$Fe/$^{56}$Fe, $^{58}$Fe/$^{56}$Fe & 2 \\
JACEE (1979+\dots+1995) & \cite{1998NuPhS..60...83T}$^{(b)}$  &  Fe, CNO, NeMgSiS, AllPArticles & 30\\
RUNJOB (1995+1996) & \cite{2001APh....16...13A}$^{(b)}$  &  H, He, Fe, CNO, Ne-S, AllParticles & 30 \\
RUNJOB (1995+1996+1997+1999)\!\!\!\!& \cite{2005ApJ...628L..41D}$^{(b)}$  &  LiBeB/CNO, subFe/Fe, CNO, AllParticles & 17\\
SOKOL (1984/03-1986/01) & \cite{1993ICRC....2...17I}$^{(b)}$  &  CNO, NeMgSiS, AllParticles & 23\\
  \multirow{2}{*}{OLIMPIYA (2005-2016)}& \multirow{2}{*}{\cite{2022AdSpR..70.2674A}}\hspace{0.5cm}\rdelim\{{2}{-2mm}\!\!\!\!\!&  Z=(70,$\cdots$,83)/(Z$\geq$70),\; Z=(55,$\cdots$,83)/(Z$\geq$55),\;(Z$\geq$88)/(Z=74-80)& \multirow{2}{*}{78} \\
  && Z$\geq$88)/(Z=74-87),\; (Z=81-87)/(Z=74-80) & \\
  ACE-CRIS (1997/12-2000/04) & \cite{2001AdSpR..27..773W}  & $^{54,55,57,58}$Fe/$^{56}$Fe, $^{57}$Co/$^{59}$Co, $^{60,61,62,64}$Ni/$^{58}$Ni & 9 \\
  ACE-CRIS (1997/12-2014/09) & \cite{2016Sci...352..677B} & $^{60}$Fe/$^{56}$Fe, $^{60}$Fe/Fe & 2 \\
  \multirow{4}{*}{ACE-CRIS (1997/12-2019/02)}& \multirow{4}{*}{\cite{2022ApJ...936...13B}}\hspace{0.5cm}\rdelim\{{4}{-2mm}\!\!\!\!\!& ($Z=28,\cdots,38$)/Fe, $^{63,65}$Cu/Cu, $^{64,66,67,68,70}$Zn/Zn, $^{67,69,71}$Ga/Ga, & \multirow{4}{*}{37} \\
                                &  & $^{70,71,72,73,74,76}$Ge/Ge, $^{73,75}$As/As, $^{(80+82)}$Se/Se, $^{79,81}$Br/Br, & \\
                                & & $^{(74+75+76+77+78)}$Se/Se, $^{(78+79+80+81+82)}$Kr/Kr, $^{(83+84+85+86)}$Kr/Kr, \\
                                & & $^{(84+85+86)}$Sr/Sr, $^{(87+88)}$Sr/Sr  & \\
  \multirow{2}{*}{AMS02 (2011/05-2018/05)} & \multirow{2}{*}{\cite{2021PhR...894....1A}}\hspace{0.5cm}\rdelim\{{2}{-2mm}\!\!\!\!\!& $Z=1,\cdots,8$,\; ($Z=2,\cdots,7$)/O,\; N/B, ($Z=3,\cdots,5$)/C, & \multirow{2}{*}{1693}\\
                                &                             & $^1$H,\; $^1$H/(H,He), $e^{-,+}$,\; $e^{-}+e^{+}$,\; $e^{+}/(e^{-}+e^{+})$&  \\
  AMS02 (2011/05-2019/10)       & \cite{2021PhRvL.126h1102A} & F, F/B, F/Si                  & 144 \\
  AMS02 (2011/05-2019/10)       & \cite{2021PhRvL.126d1104A} & Fe, Fe/(O,He,Si)              & 184 \\
  AMS02 (2011/05-2019/10)       & \cite{2021PhRvL.127b1101A} & Na, Al, (Na,Al)/Si, (Na,Al)/F & 288 \\
  AMS02 (2011/05-2021/05)    & \cite{2023PhRvL.130u1002A} & B, C, O, F, Ne, Mg, Si, S & 486 \\
  CALET (2015/10-2019/10)       & \cite{2020PhRvL.125y1102A} & C, O, C/O                     & 66  \\
  CALET (2016/01-2020/05)       & \cite{2021PhRvL.126x1101A} & Fe                            & 23  \\
  CALET (2015/10-2022/02)       & \cite{2022PhRvL.129y1103A} & B, C, B/C                     & 60  \\
  CALET (2015/10-2021/12)       & \cite{2022PhRvL.129j1102A} & H                             & 29  \\
  CALET (2015/11-2021/05)       & \cite{2022PhRvL.128m1103A} & Ni                            & 7   \\
  CALET (2015/10-2022/04)    & \cite{2023PhRvL.130q1002A} & H/He, He & 49 \\
  DAMPE (2016/01-2018/06)       & \cite{2019SciA....5.3793A} & H                             & 17  \\
  DAMPE (2016/01-2020/06)       & \cite{2021PhRvL.126t1102A} & He                            & 23  \\
  DAMPE (2016/01-2021/12)       & \cite{2022SciBu..67.2162D}\!\!\!\!& B/C, B/O                      & 26  \\
  IMP8 (1982/07-1982/12)        & \cite{1986JGR....91.2858G} &  H,He                          & 18  \\
  IMP8 (1984/08-1984/09)        & \cite{1986JGR....91.2858G} & H,He                          & 17  \\
  ISS-CREAM (2017/08-2019/02) & \cite{2022ApJ...940..107C} & H  & 13 \\
  NUCLEON (2015/07-2017/06) & \cite{2020JETPL.111..363K} & H/He& 18 \\
  NUCLEON (2015/07-2017/06) & \cite{2021BRASP..85..353T} & Ni & 19 \\
  NUCLEON (2015/07-2017/06)& \cite{2019AdSpR..64.2559G} & (Z=21-24)/Fe & 24\\
  NUCLEON-IC (2015/07-2017/06)  & \cite{2019AdSpR..64.2546G}$^{(c)}$ &  H, He, C, O, Ne, Mg, Si, Fe, AllParticles& 68 \\
  NUCLEON-KLEM (2015/07-2017/06)& \cite{2020PhLB..81135851K} & He/O, C/O& 19 \\
  PAMELA-CALO (2006/07-2014/09) & \cite{2021Univ....7..183N} & $^7$Li/$^6$Li, $^7$Be/Be, $^{10}$Be/$^9$Be & 4 \\
  PAMELA-TOF (2006/07-2014/09)  &  \cite{2021Univ....7..183N}& $^7$Li/$^6$Li, $^7$Be/Be, $^{10}$Be/$^9$Be & 8    \\
  \multirow{2}{*}{Skylab (1973/05-1974/02)}   & \multirow{2}{*}{\cite{1978ApJ...220..719S}}\hspace{0.5cm}\rdelim\{{2}{-2mm}\!\!\!\!\!& (Z=74-87)/(Z$\geq$65),\; Z$\geq$88/(Z=74-87),\; Z$\geq$65, & \multirow{2}{*}{12}\\
& & (Z$\geq$94)/(Z=74-87),\; (Z$\geq$110)/(Z=74-87) & \\
Voyager2-HET (1986/01-1987/12) & \cite{1991AaA...247..163F}$^{(b)}$  &  H, C, N, O, Ne, Mg, Si, S, Fe & 124\\
\hline
\end{tabular}
\\
$^a$ Isotopic ratios from tables were already included in \crdb{}: the new points added are extracted from the figures.\\
$^b$ Publications whose data were only partly extracted in the previous release.\\
$^c$ These KLEM data were already included in \crdb{}, but mislabelled NUCLEON.
}
\end{table*}

\subsection{Data uploaded since \crdbv{4.0}}
\label{sec:v4.1-Std}
Many data from AMS-02, CALET, DAMPE, etc. have been published since our last release. These data sets should have ideally been uploaded in \crdb{} shortly after their publication, but were only prepared for this release. We also took the opportunity of this release to upload a few old datasets that were not yet in \crdb{}. 
Rather than a detailed and cumbersome description of all these new data sets, which are listed in Table~\ref{table:new_generic}, we prefer to highlight below some of their most salient features.

To start with, the first 7 years of AMS-02 data \cite{2021PhR...894....1A}, along with other publications by the AMS collaboration \cite{2021PhRvL.126h1102A,2021PhRvL.126d1104A,2021PhRvL.127b1101A}, all uploaded in this release, now provide the {\em most comprehensive set of data from a single experiment}. These data are in the GV to TV rigidity range, and correspond to fluxes and ratios of leptons, antiprotons, and nuclei from H to Si, plus Fe. Moreover, in addition to the above AMS-02 data, we have uploaded the recent CALET \cite{2020PhRvL.125y1102A,2021PhRvL.126x1101A,2022PhRvL.129y1103A,2022PhRvL.129j1102A,2022PhRvL.128m1103A}, DAMPE \cite{2019SciA....5.3793A,2021PhRvL.126t1102A,2022SciBu..67.2162D}, ISS-CREAM \cite{2022ApJ...940..107C}, and NUCLEON \cite{2019AdSpR..64.2546G,2019AdSpR..64.2559G,2020PhLB..81135851K,2020JETPL.111..363K,2021BRASP..85..353T} data, which provide the {\em most precise set of direct measurement data in the TeV domain and above}; these data are key to investigate possible breaks and features in the spectra, and the consistency between direct and indirect measurement data. 

Some of the new data sets uploaded also explore in a unique way the composition of ultra-heavy CRs (UHCR). Indeed, recent ACE-CRIS data \cite{2022ApJ...936...13B} {\em unveil the isotopic content of CR elements $Z=30-38$}, complementing the elemental fractions measured by Tiger and SuperTiger (already in \crdb{}); a further extension to the range $41\leq Z\leq56$ should be available soon by SuperTiger \cite{2022AdSpR..70.2666W}. For even heavier (and rarer) elements, very few experiments have provided data so far. In addition to Ariel6, HEAO3-HNE, UHCRE-LDEF, and Trek data (already in \crdb{}), we added the skylab data \cite{1978ApJ...220..719S}. 

The last piece of UHCR data that we decided to add in this release are those from the OLIMPIYA experiment. The latter uses olivine crystals contained in stony-iron meteorites (pallasites) as CR detectors. At variance with satellite experiments that provide measurements of UHCR GCRs accumulated over an exposure time of a few years, the OLIMPIYA experiment provides {\em measurements of GCRs accumulated over up to hundreds of Myr}---these two complementary techniques allow to have a glimpse on the GCR time evolution. The OLIMPIYA data uploaded in this release\footnote{We stress that, owing to the DB structure and displays, we have to define a data taking period, a position in the Solar system, and an energy for these data, although it is inadequate: the former is set to the publication date, the position to 1~au, and the energy to 1.5 GeV/n (as set for the other UHCR experiments, see \cite{2020Univ....6..102M}), i.e. a value at which GCR fluxes are maximal and are likely to be responsible for most of the tracks.} are taken from  \cite{2022AdSpR..70.2674A,2022PAN....85..446A} that supersedes a previous analysis presented in \cite{2016ApJ...829..120A}\footnote{Between these two publications, several effects that could affect the relative yield of nuclei registered have been investigated and accounted for: anomalies near the meteorite edge related to the the annealing of the measured tracks \cite{2021PAN....84..849K}; fragmentation in the meteorite which explains all events in the $84\leq Z\leq 89$ range, but has no impact for the other charges \cite{2022JETP..134..523B}.}.

\subsection{Anisotropy data}
\label{sec:v4.1-Anisotropy}

Ground-based detectors with high event statistics allow the study of anisotropies in the arrival directions of CRs. Of particular interest is here the dipole anisotropy predicted by diffusion theory, that allows us to study the nearby CR source distribution and diffuse CR transport in our local magnetic environment~\cite[e.g.,][]{2017PrPNP..94..184A}.

While the true dipole anisotropy is represented by an amplitude and two phases, the data-driven reconstruction method of ground-based observatories allows only the reconstruction of the projection of the dipole vector onto the equatorial plane. Conventionally, this projection is characterised by the (projected component of the) amplitude and the phase in right ascension. These new dipole anisotropy data are indicated in the DB by two new entries, namely \code{DipoleAmplitude} and \code{DipolePhase}; we have chosen a convention where $\code{DipolePhase} \in [-180^\circ, 180^\circ]$. 

The dipole data in terms of total energy  \code{ETOT} is shown in Fig.~\ref{fig:dipoledata}. Note that the limited statistics of CR experiments in the PeV--EeV energy region has so far only yielded upper limits on the dipole anisotropy. In the DB, we indicate this by providing both the best amplitude and its upper limit as separate entries. As visible in Fig.~\ref{fig:dipoledata}, the dipole amplitude and phase data from different observatories can show strong deviations beyond statistical uncertainties. This is related to hidden (and often unquantified) systematic effects, corresponding to the partial sky coverage of experiments and reconstruction method. 
\begin{figure*}
\centering
\includegraphics[width=\linewidth]{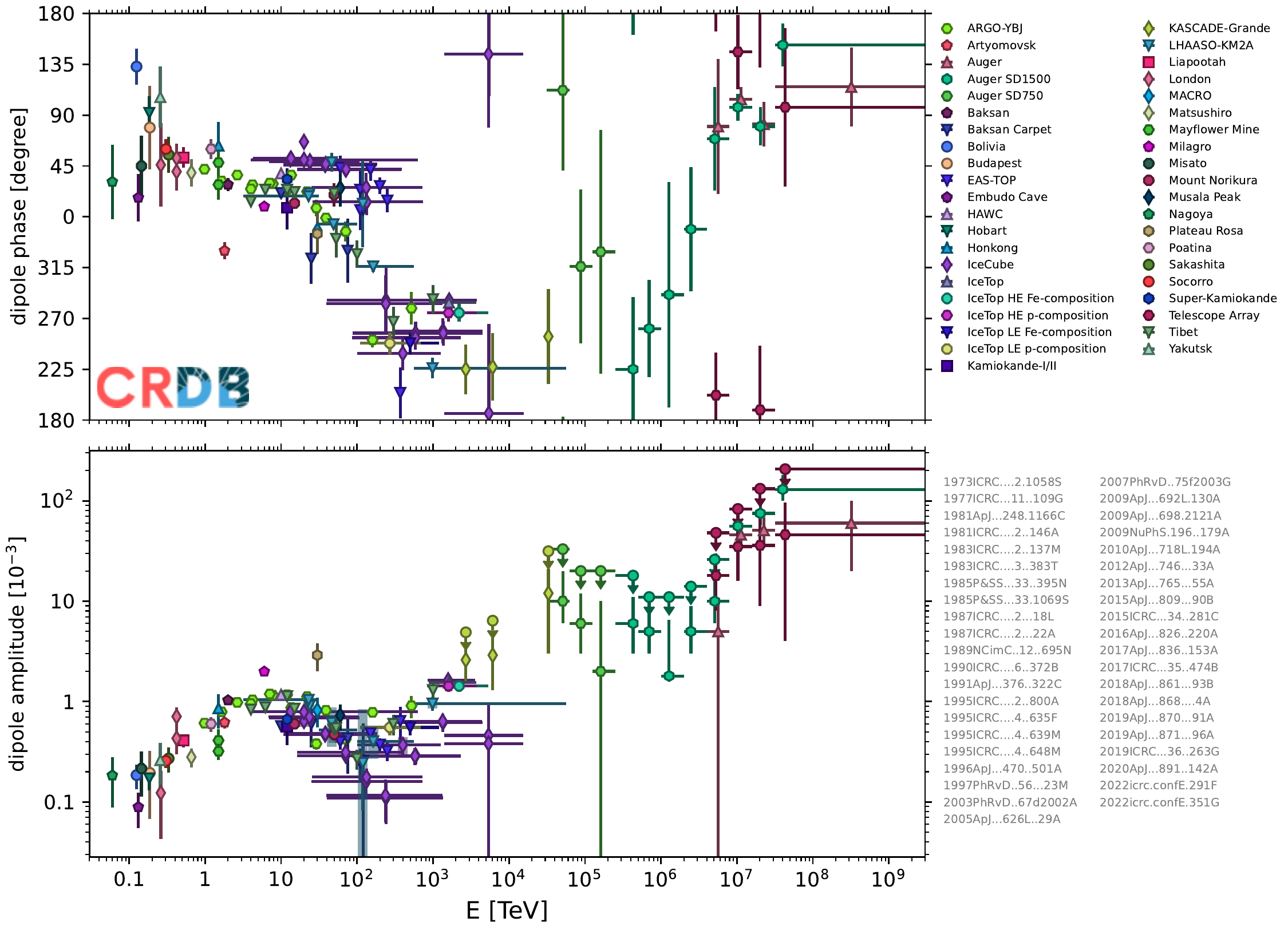}
\caption[]{Equatorial dipole amplitude and phase of the CR anisotropy inferred by various experiments (see Table~\ref{table:new_anisotropy} for references). The figure is available from the gallery notebook\footref{foot:gallery}.}\label{fig:dipoledata}
\end{figure*}

Furthermore, experimental collaborations oftentimes provide a number of updates of their anisotropy studies as the event statistics accumulate. We have chosen to include all the data publicly available, but note that the later data sets are usually meant to supersede the earlier ones. Finally, note that some of the (especially older) data have been extracted from publications, which give rather limited information on the methodology used. We have chosen to include these at face value, but recommend to exercise caution when using these data for quantitative studies. The experiments and associated references for all these data are gathered in Table~\ref{table:new_anisotropy}.

\begin{table}
\caption{Anisotropy dipole amplitude (\code{DipoleAmplitude}) and phase (\code{DipolePhase}) datasets added in \crdbv{4.1}.}
\label{table:new_anisotropy}
\centering
{
\begin{tabular}{p{6.7cm}p{0.95cm}}
\hline\hline
  Subexp \code{NAME}           & Ref.  \\
\hline
ARGO-YBJ (2008/01-2009/12)                              &  \cite{2015ApJ...809...90B} \\
ARGO-YBJ (2008/01-2012/12)                              &  \cite{2018ApJ...861...93B} \\
Artyomovsk (1981/01-1987/12)                            &  \cite{1990ICRC....6..372B} \\
Baksan (1982/07-1986/06)                                &  \cite{1987ICRC....2...22A} \\
Baksan Carpet (1980/02-1981/01)                         &  \cite{1981ICRC....2..146A} \\
Baksan Carpet (2007/01-2007/12)                         &  \cite{2009NuPhS.196..179A} \\
Bolivia (1965/01-1976/12)                               &  \cite{1985PSS...33.1069S} \\
Budapest (1958/01-1963/12)                              &  \cite{1985PSS...33..395N} \\
EAS-TOP (1990/01-1994/12) 1                             &  \cite{1996ApJ...470..501A} \\
EAS-TOP (1992/01-1994/12)                               &  \cite{1995ICRC....2..800A} \\
EAS-TOP (1992/01-1999/12)                               &  \cite{2009ApJ...692L.130A} \\
Embudo Cave (1965/01-1983/12)                           &  \cite{1985PSS...33.1069S} \\
HAWC (2015/05-2017/05)+IceCube (2011/05-2016/05)        &  \cite{2019ApJ...871...96A} \\
Hobart (1958/01-1983/12)                                &  \cite{1985PSS...33..395N} \\
Honkong (1983/11-1986/02)                               &  \cite{1987ICRC....2...18L} \\
IceCube (2007/06-2008/03)                               &  \cite{2010ApJ...718L.194A} \\
IceCube (2009/05-2015/05)                               &  \cite{2016ApJ...826..220A} \\
IceCube (2009/05-2016/05)                               &  \cite{2017ICRC...35..474B} \\
IceCube HE (2009/05-2010/05)                            &  \cite{2012ApJ...746...33A} \\
IceCube LE (2009/05-2010/05)                            &  \cite{2012ApJ...746...33A} \\
IceTop (2009/05-2015/05)                                &  \cite{2016ApJ...826..220A} \\
IceTop HE (2009/05-2012/05)                             &  \cite{2013ApJ...765...55A} \\
IceTop LE (2009/05-2012/05)                             &  \cite{2013ApJ...765...55A} \\
KASCADE-Grande (2003/01-2013/01)                        &  \cite{2019ApJ...870...91A} \\
Kamiokande-I/II (1987/12-1994/12)                       &  \cite{1997PhRvD..56...23M} \\
LHAASO-KM2A (2018/02-2018/12)                           &  \cite{2019ICRC...36..263G} \\
LHAASO-KM2A (2020/01-2020/12)                           &  \cite{2022icrc.confE.351G} \\
Liapootah (1992/01-1994/10)                             &  \cite{1995ICRC....4..639M} \\
London (1961/01-1982/12)                                &  \cite{1985PSS...33..395N} \\
London (1972/01-1978/12)                                &  \cite{1983ICRC....3..383T,1985PSS...33..395N} \\
London (1981/01-1982/12)                                &  \cite{1983ICRC....3..383T,1985PSS...33..395N} \\
MACRO (1991/11-2000/05)                                 &  \cite{2003PhRvD..67d2002A} \\
Matsushiro (1985/01-1995/12)                            &  \cite{1995ICRC....4..648M,1995ICRC....4..639M} \\
Mayflower Mine (1978/01-1981/08)                        &  \cite{1981ApJ...248.1166C} \\
Mayflower Mine (1978/01-1983/05)                        &  \cite{1991ApJ...376..322C} \\
Milagro (2000/07-2007/07)                               &  \cite{2009ApJ...698.2121A} \\
Misato (1978/01-1980/12)                                &  \cite{1985PSS...33..395N} \\
Mount Norikura (1970/08-1972/12)                        &  \cite{1973ICRC....2.1058S} \\
Mount Norikura (1975/01-1987/12)                        &  \cite{1989NCimC..12..695N} \\
Musala Peak (1968/09-1972/12)                           &  \cite{1975Natur.255..687G,1977ICRC...11..109G} \\
Nagoya (1971/01-1980/12)                                &  \cite{1985PSS...33..395N} \\
Ottawa (1977/01-1980/12)                                &  \cite{1981ICRC...10..246B} \\
Pierre Auger (2004/01-2016/08) Rayleigh          &  \cite{2018ApJ...868....4A} \\
Pierre Auger SD1500 (2004/01-2018/08) East-West  &  \cite{2020ApJ...891..142A} \\
Pierre Auger SD1500 (2004/01-2018/08) Rayleigh   &  \cite{2020ApJ...891..142A} \\
Pierre Auger SD750 (2012/01-2018/08) East-West   &  \cite{2020ApJ...891..142A} \\
Plateau Rosa (1981/05-1982/04)                          &  \cite{1983ICRC....2..137M} \\
Poatina (1987/01-1994/12)                               &  \cite{1995ICRC....4..635F} \\
Sakashita (1978/01-1980/12)                             &  \cite{1985PSS...33..395N,1990ICRC....6..361U} \\
Socorro (1968/01-1983/12)                               &  \cite{1985PSS...33.1069S} \\
Super-Kamiokande (1996/06-2001/05)                      &  \cite{2007PhRvD..75f2003G} \\
Telescope Array (2008/05-2020/05)                       &  \cite{2022icrc.confE.291F} \\
Tibet II \& III (1995/10-2010/02)                        &  \cite{2017ApJ...836..153A} \\
Tibet III (1999/11-2003/11)                             &  \cite{2005ApJ...626L..29A} \\
Yakutsk (1957/07-1969/12)                               &  \cite{1985PSS...33..395N} \\
\hline
\end{tabular}
}
\end{table}

\begin{table*}
\caption{UHECR datasets (from ground experiments) uploaded in \crdbv{4.1}.}
\label{table:new_UHECR}
\centering
{
\begin{tabular}{l l l}
\hline\hline
  Subexp \code{NAME}           & Ref.  & Qty                 \\
\hline
AGASA (1993/12-2002/12)                 & \cite{2003APh....19..447T} &  AllParticles  \\
Auger Hybrid (2004/01-2016/12) QGSJet-II-04+EPOS-LHC+SIBYLL2.3\!\!\!\!\!\!\!& \cite{2017ICRC...35..486F} & AllParticles\\
Auger (2004/01-2007/08)                                        & \cite{2008PhRvL.101f1101A}  & AllParticles  \\
Auger (2005/11-2008/05)                                        & \cite{2010PhLB..685..239A}  & AllParticles  \\
Auger SD750+SD1500 (2014/01-2018/08)                           & \cite{2021EPJC...81..966A}  & AllParticles  \\
Auger SD750 (2014/01-2018/08)                                  & \cite{2021EPJC...81..966A}  & AllParticles  \\
DICE (1994/07-1996/03)                  & \cite{2000APh....13..137S} &  AllParticles  \\
EAS-TOP (1992/01-1999/12) QGSJet        & \cite{1999APh....10....1E} &  allParticles  \\
Fly's Eye (1988/06-1992/07)             & \cite{1994ApJ...424..491B} &  AllParticles  \\
GAMMA (2003/01-2007/12) SIBYLL        & \cite{2014PhRvD..89l3003T} &  H, He,O-group, Fe-group, AllParticles  \\
GAMMA (2004/01-2006/12) SIBYLL          & \cite{2008JPhG...35k5201G} &  AllParticles  \\
Haverah Park (1974/01-1987/12) QGSJet98 & \cite{2003APh....19...47A} &  AllParticles  \\
HAWC (2016/06-2017/02) QGSJet-II-03                            & \cite{2017PhRvD..96l2001A}  & AllParticles  \\
HAWC (2018-2019) QGSJet-II-04                                  & \cite{2022icrc.confE.330M}  & AllParticles  \\
HEGRA (1987-1998) QGSJet       &\cite{2000AaA...359..682H}& AllParticles  \\
IceCube (2010/06-2013/05) SIBYLL2.1                           & \cite{2019PhRvD.100h2002A}  & H, He, O-group, Fe-group, AllParticles \\
IceCube+IceTop (2010/06-2013/05) SIBYLL2.1                           & \cite{2019PhRvD.100h2002A}  & AllParticles \\
IceTop (2016/05-2017/04) QGSJet-II-04                          & \cite{2020PhRvD.102l2001A}  & AllParticles  \\
IceTop (2010/06-2011/05) SIBYLL2.1                            & \cite{2013PhRvD..88d2004A}  & AllParticles  \\
IceTop (2016/05-2017/04) SIBYLL2.1                            & \cite{2020PhRvD.102l2001A}  & AllParticles  \\
KASCADE (1996/10-2002/01) QGSJet01                             & \cite{2005APh....24....1A}  & H, He, C-Fe-group, AllParticles                              \\
KASCADE (1996/10-2002/01) SIBYLL2.1 & \cite{2005APh....24....1A} & H, He, Si-group,\! Fe-group,\! C-Fe-group,\! C-group,\! AllParticles\!\!\!\\
KASCADE-Grande (2003/12-2011/10) SIBYLL2.3&\cite{2017ICRC...35..316A}& H, He-C-group, Si-Fe-group, AllParticles  \\
KASCADE-Grande (2003/12-2009/03) QGSJet-II-2 & \cite{2009arXiv0906.4007K} &  H, He, Si-group, Fe-group, CNO-group, AllParticles \\
KASCADE-Grande (2004/03-2010/12) QGSJet-II-3 & \cite{2011PhRvL.107q1104A} &  H-He-group, C-Fe-group, AllParticles \\
KASCADE-Grande (2004/03-2012/05) QGSJet-II-2 & \cite{2012APh....36..183A} &  AllParticles \\
KASCADE-Grande (2004/03-2012/05) QGSJet-II-2&\cite{2013PhRvD..87h1101A}& H-He-group, C-Fe-group, AllParticles  \\
Maket-ANI (1997-2004) QGSJet-II-01                             & \cite{2007APh....28...58C}  & H-He-group, C-Fe-group, AllParticles  \\
Mount Norikura (1976/01-1995/12)        & \cite{1997ICRC....4..117I} &  AllParticles  \\
TAIGA-HiScore (2017/01-2018/12)         & \cite{2022JETP..134..469A} &  AllParticles  \\
TALE (2014/06/-2016/03) QGSJet-II-03    & \cite{2018ApJ...865...74A} &  AllParticles  \\
Tibet III (2000/11-2004/10) QGSJet01    & \cite{2008ApJ...678.1165A} &  H-He-group, O-Fe-group  \\
Tibet III (2000/11-2004/10) SIBYLL2.1   & \cite{2008ApJ...678.1165A} &  O-Fe-group  \\
Yakutsk (1995/01-2008/12) QGSJet01      & \cite{2009NJPh...11f5008I} &  AllParticles  \\
\hline
\end{tabular}
}
\end{table*}

\subsection{UHECR data from \kcdc{}}
\label{sec:v4.1-UHECRs}
Considering the vast amount of academic databases and search engines for locating and accessing published scientific data, unified access to published datasets and spectra is still in the early stages. This is due to the large variety of experiments and thus the large variety of measured data. In cooperation with \crdb{}, the `KASCADE Cosmic-ray Data Centre'  (\kcdc{}) is taking a step towards simplification, by embedding the UHECR data from \kcdc{}, i.e. data from extensive air shower experiments,  into \crdb{}. The advantage of such an extensive collection of UHECR data is that data from other experiments can be obtained relatively quickly.
\kcdc{} is already a demonstrator and partner of PUNCH4NFDI\footnote{\url{https://www.punch4nfdi.de/}}, the consortium of particle, astroparticle, astro-, hadron and nuclear physics within the German National Research Data Infrastrucutre, NFDI, which is aimed to unify the methodical approach of open data in this field.

The \kcdc{} is a web-based interface where initially the scientific data from the completed air-shower experiment KASCADE-Grande was made available for the astroparticle community as well as for the interested public
Besides a DataShop to download the reconstructed data of KASCADE-Grande and the meta-data, \kcdc{} offers more than 100 cosmic ray spectra from about 25 different ground-based high-energy CR experiments published between 1984 and 2021 for download. The data sets available cover an energy range from about $10^{12}$~eV to more than $10^{20}$~eV for all-particle spectra (keyword \code{AllParticles} in \crdb{}) as well as for mass groups like p, He up to Fe or heavy and light respectively, derived from the unfolding procedure for different high-energy interaction models like QGSJet, EPOS and also SIBYLL, mostly embedded in the CORSIKA simulation pac\-kage: CORSIKA\footnote{\url{ https://www.iap.kit.edu/corsika/}} (COsmic Ray event SImulation for KAscade) has been written especially for KASCADE and extended since then to become the world’s standard simulation package in the field of cosmic ray air shower simulations. 

While the KASCADE-Grande experimental data in \kcdc{} are accessible also via an API, the spectra points and metadata, stored in a postgres database, can only be selected and displayed on the website after registration. Thus, a partnership with \crdb{} was set up with the aim of creating a basis for this data exchange and to provide the community with a common interface to this merged spectra data.
The \kcdc{} data sets are now being reformatted to meet the requirements of \crdb{}, to supplement its very extensive content with data from ground-based air shower experiments. The spectra uploaded on \crdb{} at the time of this release are listed in Table~\ref{table:new_UHECR}; they represent about $\sim 50\%$ of the full data being prepared, and a sample of these data can be seen in Fig.~\ref{fig:gallery_1}. To match the requirements of UHECR measurements, the data quantity list \code{DATA\_QTY} had to be extended by two more groups, the \code{He-C-group} and the \code{Si-Fe-group}.

To find out more about the real meaning of the particle spectra like helium, oxygen and so on, their mixtures as well as the mixtures of different high-energy interaction models, users should refer to the original papers.

\subsection{Upper limit on high-energy $\bar{p}/p$}
\label{sec:v4.1-ULpbarp}

With the angular resolution of ground cosmic-ray detectors reaching below the degree level in the 90's, it became possible to observe a deficit of events from the direction of the Moon or the Sun ($\sim 0.5^\circ$): the Moon or Sun shadow technique was used first to calibrate their angular resolution and pointing accuracy. Actually, the position of the shadow is offset from the true location of the blocking bodies owing to the deflection of cosmic rays in the geomagnetic field, with the shadow shifted westward (resp. eastward) for positively (resp. negatively) charged particles. This allowed several experiments to set upper limits on the $\bar{p}/p$ ratio above TeV energies \cite{1995ICRC....3...84A,2003APh....20..145A,2005APh....23..411L,2007APh....28..137T,2012PhRvD..85b2002B,2018PhRvD..97j2005A}.

These upper limits were added in \crdb{}, along with the older upper limits obtained from the observed charged ratio of muons \cite{1985AaA...149....1S}. These new datasets are shown in Fig.~\ref{fig:pbarp_UL} and listed in Table~\ref{table:new_pbarp}.
\begin{figure}
\centering
\includegraphics[width=\linewidth]{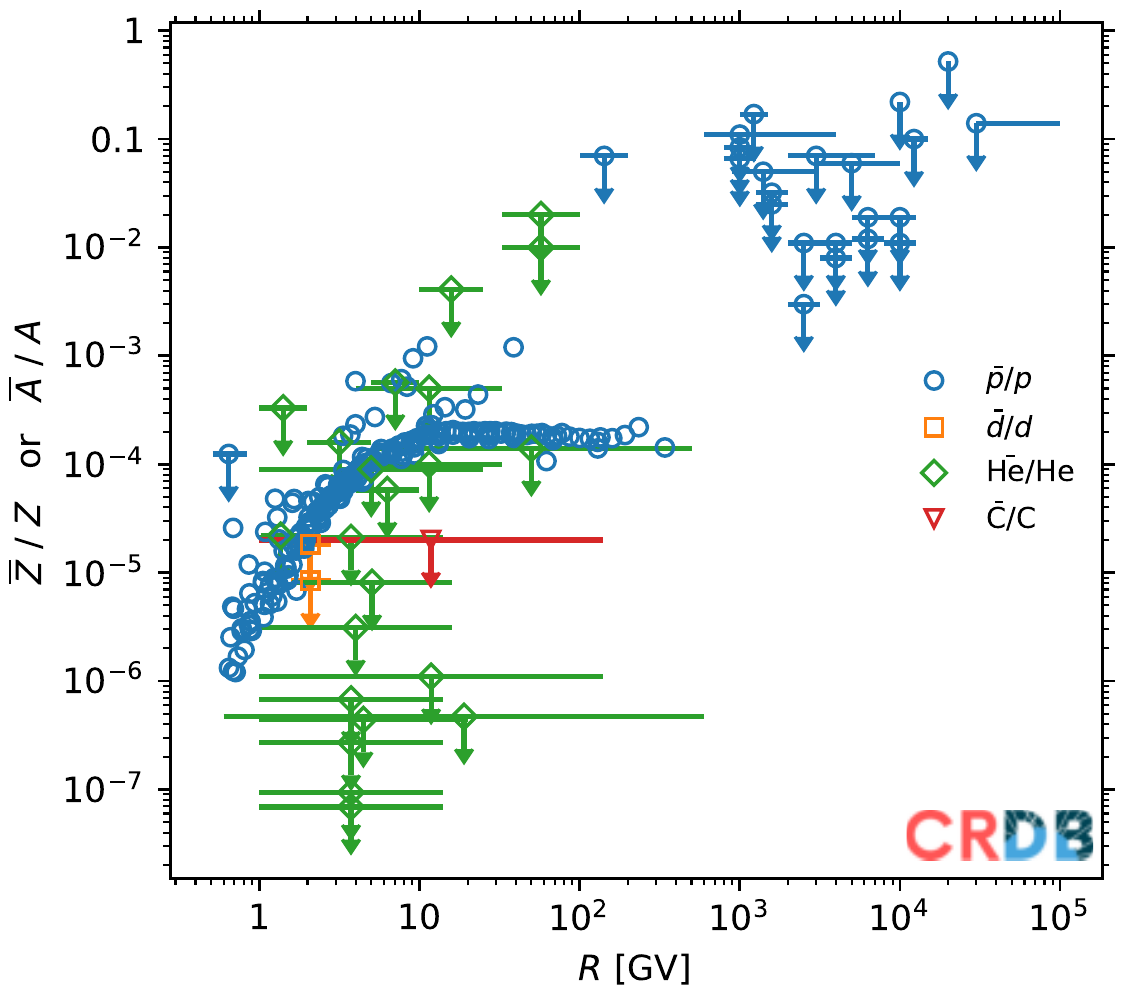}
\caption[]{$\bar{A}/A$ and  $\bar{Z}/Z$ ratios in \crdb{}; the very few data points for upper limits on $(Z\leq-2)/(Z\geq2)$, $(Z\leq-3)/(Z\geq3)$ and $(Z\leq-6)/(Z\geq6)$ are not shown. The  orange crosses with downward arrows correspond to the new $\bar{p}/{p}$ upper limits at high energy added in this release (see Table~\ref{table:new_pbarp}). The figure is available from the gallery notebook\footref{foot:gallery}.}\label{fig:pbarp_UL}
\end{figure}

\begin{table}
\caption{$\bar{p}/p$ upper limits added in \crdbv{4.1}.}
\label{table:new_pbarp}
\centering
{
\begin{tabular}{l l}
\hline\hline
  Subexp \code{NAME}           & Ref.  \\
\hline
  ARGO-YBJ (2006/07-2009/11)   & \cite{2012PhRvD..85b2002B} \\
  HAWC (2014/11-2017/08)       & \cite{2018PhRvD..97j2005A} \\
  L3+C (1999/04-2000/11)       & \cite{2005APh....23..411L} \\
  MACRO (1989/02-2000/12)      & \cite{2003APh....20..145A} \\
  Muon data (1971-1981)        & \cite{1985AaA...149....1S} \\
  Tibet I (1990/06-1993/09)    & \cite{1995ICRC....3...84A} \\
  Tibet III (1999/11-2004/12)  & \cite{2007APh....28..137T} \\
\hline
\end{tabular}
}
\end{table}

\subsection{LEE, AESOP, and AESOP-Lite balloon flights}
\label{sec:v4.1-LEE-AESOP}

From 1968 to 2011, the LEE (Low Energy Electrons) balloon-borne instrument~\cite{1970NucIM..85...93H} was launched over 35 times. LEE provided the longest series of CR electron measurements ($e^-+e^+$) over a  time period that covers about four solar cycles. This data is particularly relevant to the study of the solar modulation of electrons with energies up to about 20 GeV. In \crdbv{4.1}, we reorganized the existing LEE data from 1968 to 1994. Data points taken from figures were updated with the actual values when private communication with the authors was possible. Data post-1994 were also added to the database. Indeed, the spectra for the years 1997 to 2000 were never fully published. However, flight data were analyzed using the same method as that outlined in~\cite{1975JGR....80.1701F}, and the spectrum values at 1.2~GeV only were published in~\cite{2004JGRA..109.7107C}. The full spectra for these years were provided by the authors (Paul Evenson, 2023) and uploaded in \crdb{}.  These data are shown in the top panel of Fig.~\ref{fig:time_series_NMphi} along with other measurements from experiments at similar energies. We also show on this plot times series of He (second panel), NM count rates (third panel), and Solar modulation values calculated from these count rates (fourth panel).

From 1994 to 2011, the AESOP (Anti-Electron Sub Orbital Payload) balloon-borne instrument~\cite{1996ApJ...464..507C} flew at multiple occasions with the primary objective to study the charge-sign dependence of the solar modulation of electrons from a few hundreds MeV to a few GeV. In \crdbv{4.1}, we reorganized the existing AESOP $e^+/(e^-+e^+)$ data and updated the 1994 flight (private communication with the author John Clem, 2023).

\begin{table}
\caption{Lepton data from the LEE, AESOP, and AESOP-Lite balloons flown over a 50-year time period, re-organised, corrected, and with a few new data sets added in \crdbv{4.1}.}
\label{table:LEE-AESOP}
\centering
{
\begin{tabular}{lllr}
\hline\hline
  Subexp \code{NAME}           & Ref.                   & Qty       &$N_{\rm data}$\\
\hline
   LEE (1968/06-1968/07)        & \cite{1975JGR....80.1701F} & $e^-+e^+$ & 16     \\
  LEE (1969/06-1969/07)        & \cite{1975JGR....80.1701F} & $e^-+e^+$ & 18     \\
  LEE (1970/06-1970/07)        & \cite{1975JGR....80.1701F} & $e^-+e^+$ & 17     \\
  LEE (1971/06-1971/07)        & \cite{1975JGR....80.1701F} & $e^-+e^+$ & 18     \\
  LEE (1972/07)                & \cite{1975JGR....80.1701F} & $e^-+e^+$ & 16     \\
  LEE (1973/07)                & \cite{1975ICRC....3.1000C} & $e^-+e^+$ & 30     \\
  LEE (1974/07)                & \cite{1975ICRC....3.1000C} & $e^-+e^+$ & 16     \\
  LEE (1975/07)                & \cite{1977ICRC...11..203C} & $e^-+e^+$ & 10     \\
  LEE (1977/07)                & \cite{1983ApJ...275L..15E} & $e^-+e^+$ & 12     \\
  LEE (1979/08-1979/09)        & \cite{1984JGR....89.2647E} & $e^-+e^+$ & 8      \\
  LEE (1982/08)                & \cite{1984JGR....89.2647E} & $e^-+e^+$ & 7      \\
  LEE (1984/09)                & \cite{1986JGR....91.2858G} & $e^-+e^+$ & 10     \\
  LEE (1987/08)                & \cite{1995JGR...100.7873E} & $e^-+e^+$ & 8      \\
  LEE (1990/08)                & \cite{1995JGR...100.7873E} & $e^-+e^+$ & 6      \\
  LEE (1992/08)                & \cite{1995JGR...100.7873E} & $e^-+e^+$ & 7      \\
  LEE (1994/08)                & \cite{1995JGR...100.7873E} & $e^-+e^+$ & 8      \\
  LEE (1997/09)                & This paper\!\!\!\!& $e^-+e^+$ & 9      \\
  LEE (1998/08-1998/09)        & This paper\!\!\!\!& $e^-+e^+$ & 8      \\
  LEE (1999/08)                & This paper\!\!\!\!& $e^-+e^+$ & 9      \\
  LEE (2000/08)                & This paper\!\!\!\!& $e^-+e^+$ & 8      \\
  LEE (2002/08)                & \cite{2004JGRA..109.7107C} & $e^-+e^+$ & 15     \\
  LEE (2009/05)                & \cite{2020ApJ...903...21M} & $e^-+e^+$ & 15     \\
  LEE (2011/05)                & \cite{2020ApJ...903...21M} & $e^-+e^+$ & 15     \\[1mm]
  AESOP (1994/08)              & \cite{1996ApJ...464..507C} & $e^+/(e^-+e^+)$ & 1      \\
  AESOP (1997/09+1998/08)\!\!\!\!& \cite{2000JGR...10523099C} & $e^+/(e^-+e^+)$ & 4      \\
  AESOP (1999/08)              & \cite{2002ApJ...568..216C} & $e^+/(e^-+e^+)$ & 6      \\
  AESOP (2000/08)              & \cite{2002ApJ...568..216C} & $e^+/(e^-+e^+)$ & 3      \\
  AESOP (2002/08)              & \cite{2004JGRA..109.7107C} & $e^+/(e^-+e^+)$ & 3      \\
  AESOP (2006/06)              & \cite{2009JGRA..11410108C} & $e^+/(e^-+e^+)$ & 4      \\
  AESOP-Lite (2018/05)         & \cite{2020ApJ...903...21M} & $e^+/(e^-+e^+)$, $e^{-/+}$\!\!\!\!& 27     \\
\hline
\end{tabular}
}
\end{table}
The  AESOP-Lite apparatus is the successor of LEE and AESOP.  Its primary objectives are to search for the origin of low-energy electrons in the electron spectrum between 20-300~MeV, and to  provide a baseline electron spectrum at 1 au for the measurements of the Voyager probes currently transmitting data from outside the heliosphere. The $e^-$, $e^+$, and $e^+/(e^-+e^+)$ data from the AESOP-Lite's maiden flight from Sweden in 2018~\cite{2020ApJ...903...21M} were added to \crdb{}; future data will be added too.

The metadata of all these balloon flights were updated using information from the original publications. When not available, the information from the stratospheric balloon flight catalogue {\tt StratoCat}\footnote{\url{https://stratocat.com.ar/indexe.html}} was used. The list of the balloon flight names as encoded in \crdb{} along with the associated publications are listed in Table~\ref{table:LEE-AESOP}.

\begin{figure*}
\centering
\includegraphics[width=0.95\linewidth]{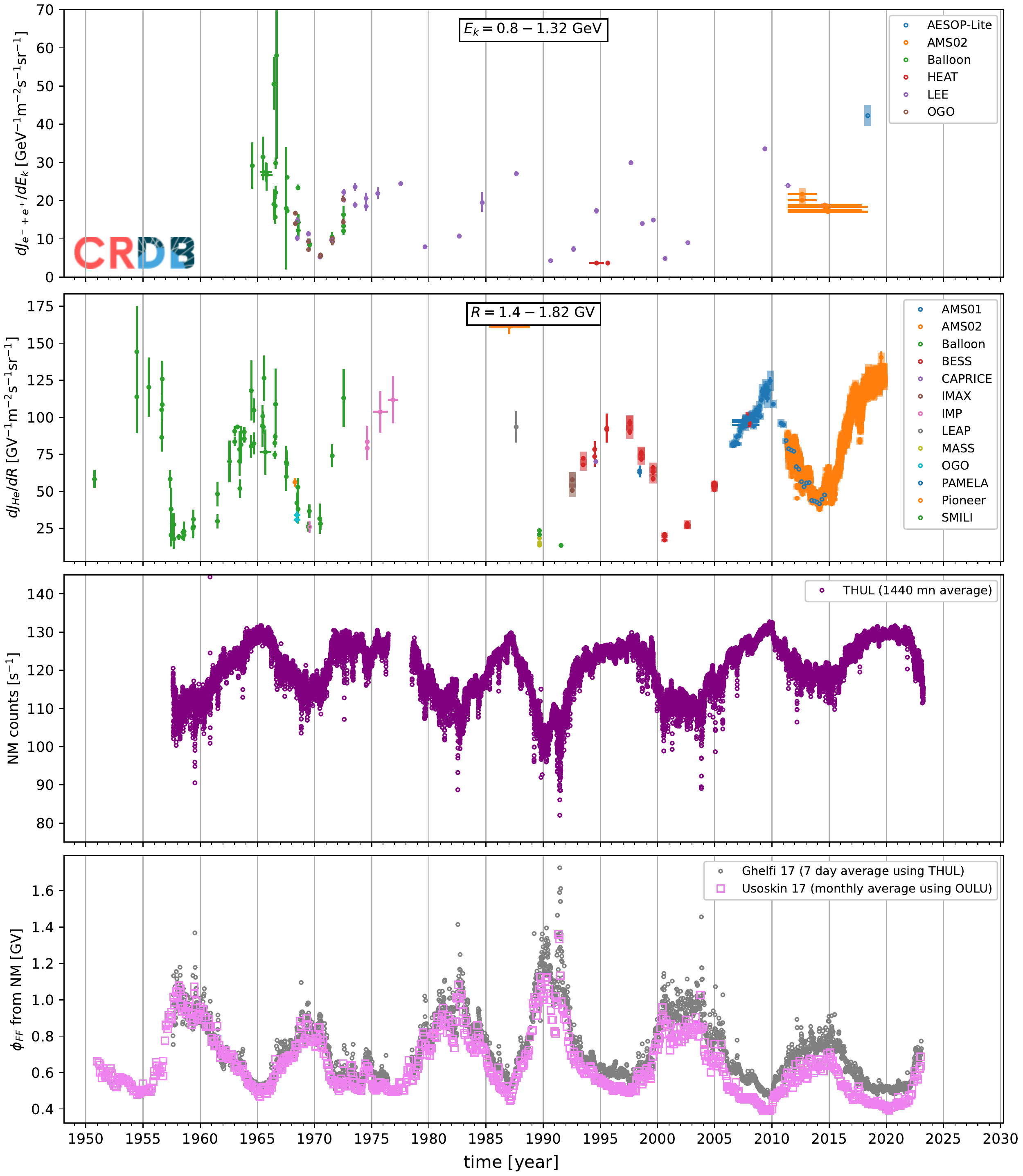}
\caption[]{{\em First and second panels}: GCR fluxes of low-energy $e^-+e^+$ and He over the last 70 years, illustrating the 11 years Solar cycle (`balloons' in the legend refers to unnamed balloons). {\em Third panel:} NM count rate from the Thule NM station retrieved from the \code{NEST} NMDB interface at \url{https://www.nmdb.eu/nest/help.php\#helptres}. {\em Bottom panel:}: Solar modulation level reconstructed from NM data \cite[e.g.,][]{2015AdSpR..55..363M}, as retrieved from \crdb{}'s {\em Solar Modulation} REST interface and whose values are based on \cite{2016A&A...591A..94G,2017A&A...605C...2G,2017AdSpR..60..833G}, or as retrieved from \url{https://cosmicrays.oulu.fi/phi/} \cite{2017JGRA..122.3875U} . The figure is available from the gallery notebook\footref{foot:gallery}.}\label{fig:time_series_NMphi}
\end{figure*}

\subsection{AMS-02 and PAMELA time series}
\label{sec:v4.1-time-series}

In \crdb{} previous releases, a few time series were already included: yearly averaged (1994-2014) proton fluxes from EPHIN \cite{2016SoPh..291..965K}, monthly or Carrington rotation average (2006-2014) proton fluxes from PAMELA \cite{2013ApJ...765...91A,2015ApJ...810..142A}, and 6 month average (2006-2009) electron fluxes from PAMELA \cite{2018ApJ...854L...2M}. 

Thanks to its large acceptance and high statistics, AMS-02 was able, for the first time, to provide daily averaged fluxes of H, He, and He/H from 2011 to 2019 \cite{2018PhRvL.121e1101A,2022PhRvL.128w1102A}, and $e^-$ from 2011 to 2021 \cite{2023PhRvL.130p1001A}: these data are now the dominant body of data in \crdb{}, with about 200\,000 data points over $\sim 3000$ days.

We also added the recently published He time series of PAMELA from 2006 to 2013. Owing to its smaller acceptance and statistics, the data were averaged over one Carrington rotation ($\sim 1$~month) in the first three years \cite{2020ApJ...893..145M}, and over three Carrington rotations later because of a {\em random failure
of a few front-end chips in the tracking system [...] particularly significant after 2009} \cite{2022ApJ...925L..24M}; this corresponds to $\sim 3000$ new data points in \crdb{} (in $E_{k/n}$ and $R$), as retrieved from the `CRDB@ASI' database\footnote{\url{https://tools.ssdc.asi.it/CosmicRays/}} \cite{2017ICRC...35.1073D}. We also added a few positron fraction data points taken from three different time periods \cite{2016PhRvL.116x1105A}: the latter paper also provides 3-month averages (2006-2016) of the $e^+/e^-$ ratio, but normalised to the unspecified 2006 value, so we did not add them in \crdb{}.

\begin{table}
\caption{AMS-02 and PAMELA time series added in \crdbv{4.1}.}
\label{table:timeseries}
\centering
{
\begin{tabular}{lllr}
\hline\hline
  Subexp \code{NAME}           & Ref.                   & Qty       &$N_{\rm data}$\\
\hline
  \multicolumn{4}{c}{\em AMS-02 (daily average)}               \\
  AMS02 (2011/05/20 to 2019/10/29) & \cite{2021PhRvL.127A1102A} &  H        & 83757  \\
  AMS02 (2011/05/20 to 2019/10/29) & \cite{2022PhRvL.128w1102A} &  He, He/H\!\!\!\!\!\!\!\!& 72879  \\
  AMS02 (2011/05/20 to 2021/11/02)        & \cite{2023PhRvL.130p1001A} & $e^-$ & 32985 \\[2mm]
  \multicolumn{4}{c}{\em PAMELA (average over Carrington rotations)}               \\
  PAMELA (2006/07-2009/12)  \rdelim\}{3}{-2mm}&                               &                 &    \\
  PAMELA (2011/05-2013/11) & \cite{2016PhRvL.116x1105A} & $e^+/(e^-+e^+)$\!\!\!\!\!\!\!\!\!\!& 15 \\
  PAMELA (2015/01-2015/12) &                               &                 &    \\[1mm]
  PAMELA (2006/07 to 2009/12)  & \cite{2020ApJ...893..145M} &  He       & 2322   \\
  PAMELA (2010/01 to 2013/09)  & \cite{2022ApJ...925L..24M} &  He       & 1026   \\
\hline
\end{tabular}
}
\end{table}

To better visualise these data, we added a new query option in the web interface to plot data as a function of time (instead of energy). The direct benefit is to enable showing the evolution of data from similar energy bands over long time periods. This is illustrated with Fig.~\ref{fig:time_series}, available from the gallery notebook\footref{foot:gallery}.
\begin{figure*}
\centering
\includegraphics[width=\linewidth]{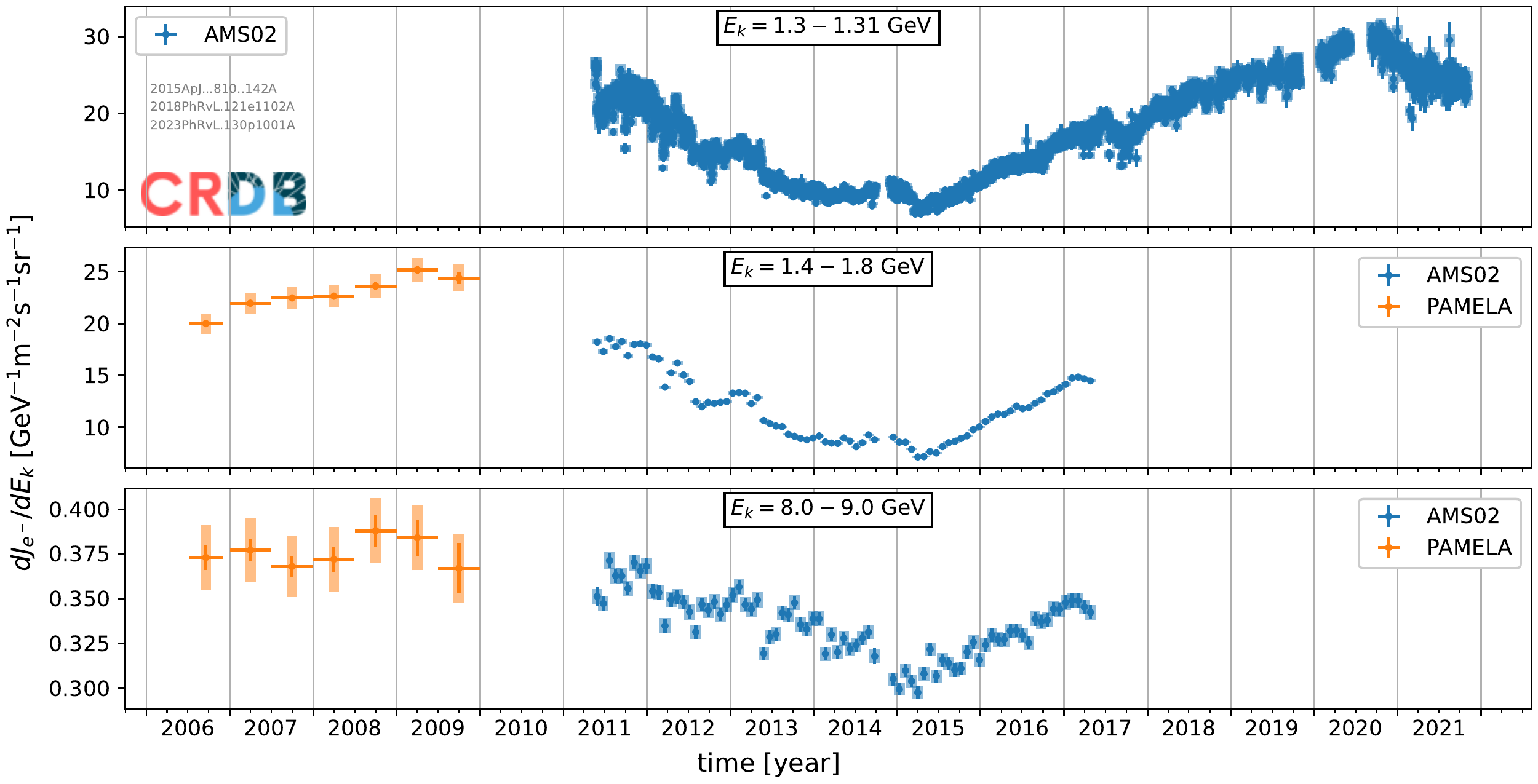}
\caption[]{Electron time series from AMS-02 \cite{2018PhRvL.121e1101A,2023PhRvL.130p1001A} and PAMELA \cite{2015ApJ...810..142A}, see Table~\ref{table:new_anisotropy}. The figure is available from the gallery notebook\footref{foot:gallery}.}\label{fig:time_series}
\end{figure*}

\section{Conclusions and future releases}
\label{sec:concl}
We have presented in this paper \crdbv{4.1}, an update of the CR database hosted at LPSC. On the technical side, this update involved a migration of \crdb{} server and a slight simplification of the DB structure. On the code side, a few minor bugs have been fixed, the queried data can now be returned in a more complete \code{csv} format (which includes all meta-data), and we fixed a missing combination rule for the data. On the web interface side, we added a new plotting capability to display CRs as a function of time, and added two new tabs: one lists all caveats related to the preparation of the data uploaded in \crdb{} and to the (sometimes approximate) transformation rules made on the queried data; the other provides a gallery of plots advertising and illustrating the diversity of \crdb{} data. Actually, this gallery and many other plots can be generated from our new public python \crdb{} library, and notebook examples are provided in the git page\footref{foot:gallery}.

On the content side, we enlarged the scope and content of \crdb{}, with the addition of dipole anisotropy data, high-energy upper limits on $\bar{p}$, a large number of UHECR datasets, and also time series data. The latter include recently released AMS-02 daily and PAMELA monthly data, but also yearly data from LEE, AESOP or AESOP-Lite balloons taken over a 50 year period. We also updated \crdb{} data with all the GCR data published in the last three years, also adding a couple of older data that had slipped our attention until now.

The path to future developments is not very clear and also depends on the feedback from the community. Indeed, \crdb{} now accounts for most galactic and extragalactic CR data, in terms of quantities that can be cast as 1D data vectors (as opposed to skymaps or higher-dimension datacubes). Missing datasets should consist mostly of old time series from satellite experiments, which are both difficult to track and retrieve from the publications: owners and authors of such datasets are welcome to get in touch with us. If need be, other quantities related to UHECR data could also be added in the future, like $\langle \ln A\rangle$. In any case, looking at present and future high-precision CR data, we stress that the current format to store uncertainties in \crdb{} is already limited and should probably be improved at some time in the future. Indeed, data from the last generation of CR detectors already come with broken-down contributions from various systematics, whereas only the total systematics can be stored in \crdb{}. This issue will worsen when covariance matrix of uncertainties will start to be released as well (as is already the case for instance for the most recent Pierre Auger data).

The \crdb{} team will continue uploading newly published CR data, but we also encourage collaborations to prepare their data (\crdb{} submission format) if they wish them to quickly be distributed via \crdb{}.
Comments, questions, suggestions, and corrections on are welcome and are to be sent at \href{mailto:crdb@lpsc.in2p3.fr}{\tt crdb@lpsc.in2p3.fr}.

\section*{Acknowledgements}
We warmly thank the continuous support and feedback from many of our colleagues, who point out typos and mismatches in \crdb{}. We also thank the AMS-02 collaboration for providing their data as \code{csv} tables (\url{https://ams02.space/publications}), which greatly eases the preparation and upload of these data in \crdb{}. This research has made use of NASA’s Astrophysics Data System Bibliographic Services. 
We acknowledge the NMDB database (\url{www.nmdb.eu}), founded under the European Union's FP7 programme (contract no. 213007) for providing data; NM data from Oulu are provided by the Sodankyla Geophysical Observatory (see also \url{https://cosmicrays.oulu.fi/readme.html}) and those from Thule by the University of Delaware Department of Physics and Astronomy and the Bartol Research Institute.

\section*{Funding}
This work was partially supported by NASA award 80NSSC\-19K0746 and partially by funding from Deutsche Forschungsgemeinschaft (DFG, German Research Foundation) -- project no.\ 449728698.

\section*{Data Availability Statement}
The authors declare that all the data underlying this article are available from \crdb{} at {\url{https://lpsc.in2p3.fr/crdb}}.

\appendix

\section{Convention for energy units}
\label{app:Eunit}

When comparing spectra of different nuclei, a convenient energy variable is energy per nucleon. 
It has become customary to indicate the use of energy per nucleon by employing non-standard units, e.g. GeV/n, which is to be contrasted to simply GeV for energy. 
After thoroughly discussing the issue, we chose to break with tradition for this paper and for the revision of the \crdb{}. 
Instead we use the energy unit GeV throughout and indicate by other means (by using different quantity symbols and via textual description) whether an energy is given per nucleon or given per particle.

The original motivation for using both GeV and GeV/n was to distinguish between energy per particle and energy per nucleon, but this distinction is redundant if a distinction via a quantity symbol (for example, $E_k$ and $E_{k/n}$) and a textual description of the quantity is also given. The latter is anyway required, since the meaning of the non-standard unit GeV/n is not universally understood, since it does not adhere to the International System of Units (SI) \cite{si-2022}. In the SI, units are mathematical entities that must be constants and combinations of SI base units: the unit of a quantity that describes an energy per nucleon is still just the energy unit.

In summary, we believe that dropping the use of GeV/n in \crdb{} will not harm the clarity of our presentation, and by following the SI, it will simplify communicating with other fields.

\bibliographystyle{spphys}
\bibliography{crdb}
\end{document}